\documentclass[manuscript]{acmart}

\usepackage{multicol}
\usepackage{tikz}
\usepackage{wasysym,booktabs,multirow,array,graphicx}
\usetikzlibrary{calc, positioning, arrows.meta, shapes.multipart, patterns}
\usepackage{xcolor}

\AtBeginDocument{%
  }

\setcopyright{acmlicensed}
\copyrightyear{2018}
\acmYear{2018}
\acmDOI{XXXXXXX.XXXXXXX}
\acmConference[Conference acronym 'XX]{Make sure to enter the correct
  conference title from your rights confirmation email}{June 03--05,
  2018}{Woodstock, NY}
\acmISBN{978-1-4503-XXXX-X/2018/06}

\begin{document}

\title{Why Johnny Can't Vibe: The Competence Paradox of AI-Assisted Software Development}

\author{Sabbir Bin Abdul Latif}
\orcid{0009-0001-1174-5783}
\affiliation{%
  \department{Institute of Information Technology}
  \institution{University of Dhaka}
  \city{Dhaka}
  \country{Bangladesh}
}
\email{ms132011@iit.du.ac.bd}

\author{Md Raihan}
\orcid{0009-0002-6711-4929}
\affiliation{%
  \department{Computer Science and Engineering}
  \institution{BRAC University}
  \city{Dhaka}
  \country{Bangladesh}
}
\email{md.raihan@g.bracu.ac.bd}

\author{Md. Mahfuzul Ahsan}
\orcid{0009-0000-6765-6379}
\affiliation{%
  \department{Computer Science and Engineering}
  \institution{BRAC University}
  \city{Dhaka}
  \country{Bangladesh}
}
\email{md.mahfuzul.ahsan@g.bracu.ac.bd}

\author{Md. Fakhruddin Gazzali}
\orcid{0000-0001-5609-5852}
\affiliation{%
  \department{Computer Science and Engineering}
  \institution{BRAC University}
  \city{Dhaka}
  \country{Bangladesh}
}
\email{fakhruddin.gazzali@bracu.ac.bd}

\author{Jannatun Noor}
\orcid{0000-0001-9669-151X}
\affiliation{%
  \institution{Computing for Sustainability and Social Good (C2SG) Research Group}
  \department{Computer Science and Engineering}
  \institution{United International University}
  \city{Dhaka}
  \country{Bangladesh}
}
\email{jannatun@cse.uiu.ac.bd}

\renewcommand{\shortauthors}{Latif et al.}

\begin{abstract}
  Large Language Models generate code, but not the responsibility that writing it carried. The work AI-assisted development was meant to remove does not disappear, it changes form. We traced how in a mixed-methods study of Bangladeshi developers, a 116-respondent survey plus contextual inquiry and interviews with 14 developers using their own tools. Developers moved planning into files the agent reads automatically, drew deliberate delegation boundaries, treated token cost as an active constraint, and described verification as absorbing much of what generation saved. Comprehension tracked authorship or shared convention, not seniority. Careful reading tracked less automation bias and fewer accepted bugs. We name this the competence paradox of vibe coding. A tool marketed as flattening expertise instead rewards a second, undocumented expertise in operating it, most visible in instruction files teams write but never validate. We turn these into design implications and a proof-of-concept skill generator, Better Skills.
\end{abstract}



\begin{teaserfigure}
    \centering
    \begin{tikzpicture}
        \node[
            draw=blue!20,
            dashed,
            rounded corners=8pt,
            line width=0.65pt,
            inner sep=4pt
        ] {
            \includegraphics[width=0.97\textwidth]{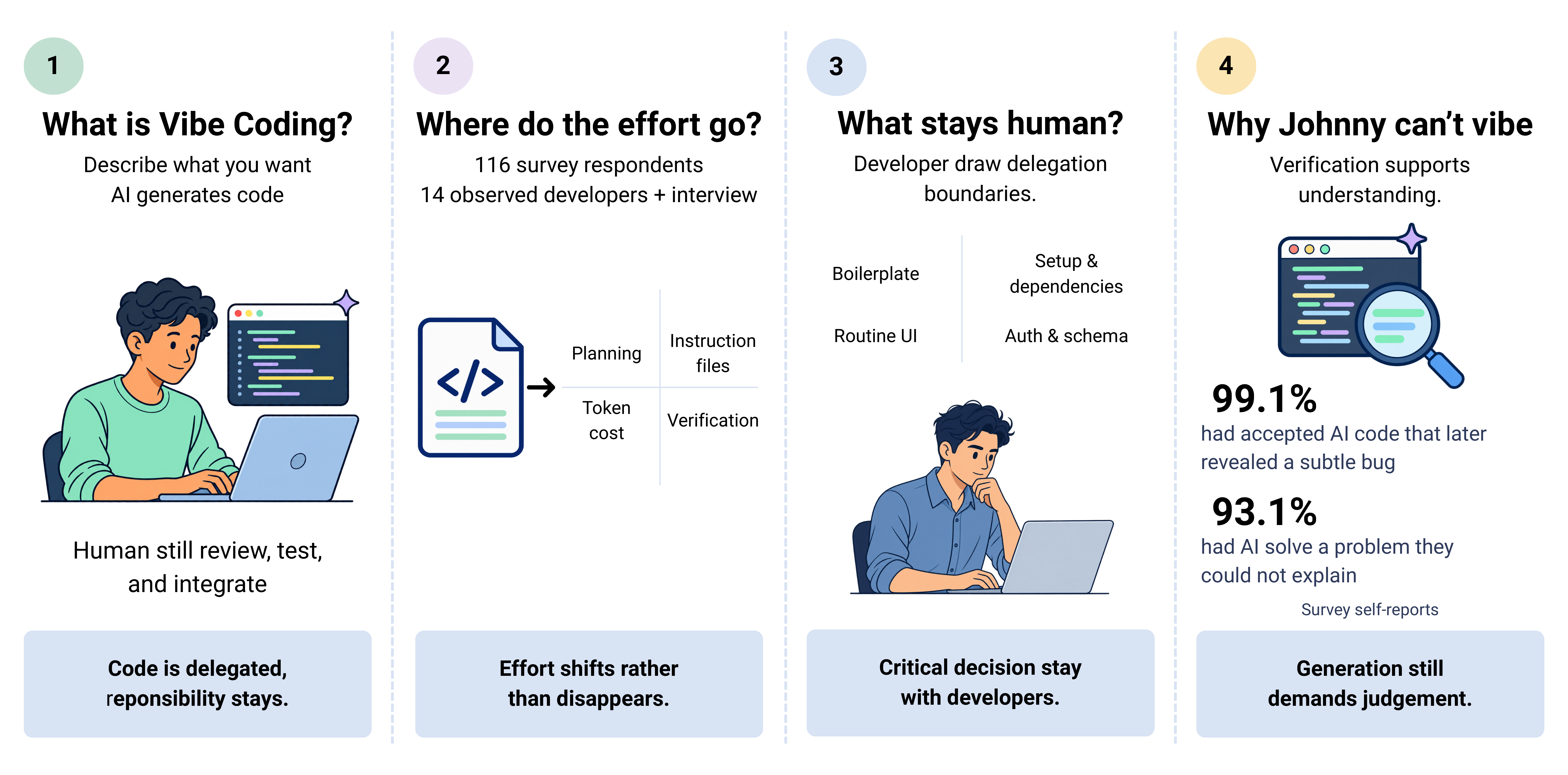}
        };
    \end{tikzpicture}
    \caption{Vibe coding shifts rather than eliminates developer effort. Across our mixed-methods study of 116 survey respondents and 14 observed developers, developers delegated code generation while retaining critical decisions, taking on new planning and verification work, and sometimes accepting AI-generated code they could not fully verify or explain.}
    \Description{A four-panel horizontal infographic separated by dashed vertical lines. Panel 1, titled "What is Vibe Coding?", shows a developer at a laptop beside an AI-generated code window and states that humans still review, test, and integrate the code. Panel 2, titled "Where does the effort go?", shows a code-document icon feeding four labeled areas: planning, instruction files, token cost, and verification; it identifies 116 survey respondents and 14 observed developers with interviews. Panel 3, titled "What stays human?", contrasts delegated boilerplate and routine user-interface work with retained setup and dependencies, authentication, and schema decisions, above an illustration of a developer thinking at a laptop. Panel 4, titled "Why Johnny can't vibe", shows a code window under a magnifying glass and reports that 99.1 percent had accepted AI code that later revealed a subtle bug and 93.1 percent had AI solve a problem they could not explain. Pale green, lavender, blue, and yellow numbered circles label the four panels, and each panel ends with a pale-blue summary box.}
    \label{fig:vibecoding-teaser}
\end{teaserfigure}


\maketitle

\definecolor{paperBlue}{HTML}{4066A3}
\definecolor{paperMid}{HTML}{7895BE}
\definecolor{paperLight}{HTML}{B7C6DD}
\definecolor{paperPale}{HTML}{D8E4F5}
\definecolor{paperWash}{HTML}{F0F4FA}
\definecolor{paperRule}{HTML}{C6D4E8}
\definecolor{paperInk}{HTML}{303E53}
\definecolor{paperMuted}{HTML}{596579}
\definecolor{paperNegative}{HTML}{977A91}
\tikzset{every picture/.append style={text=paperInk}}
\definecolor{planninginstructionlabor}{HTML}{4066A3}
\definecolor{costmodeleconomics}{HTML}{7895BE}
\definecolor{verificationsafetyburden}{HTML}{4066A3}
\definecolor{ownershipcomprehension}{HTML}{7895BE}
\definecolor{expertisegatekeeping}{HTML}{4066A3}

\newcommand{\sunburstlabel}[5]{%
    \pgfmathparse{(#1>90) && (#1<270) ? #1-180 : #1}%
    \let\labelrot\pgfmathresult
    \node[
        align=center,
        font=#4,
        text width=#3,
        inner sep=0pt,
        outer sep=0pt,
        rotate=\labelrot
    ]
    at (#1:#2)
    {#5};
}

\section{Introduction} \label{sec:introduction}
Software engineering has always been organized around packages of human knowledge. Architecture, design patterns, project management, and quality assurance are ways of dividing up a system that no single person can hold in their head. Each package carries responsibility as well as knowledge; whoever owns the design answers for it when it fails.
Large Language Models are now inserted directly into this structure, and they break that pairing. They generate code, tests, and scaffolding at a speed no human can match, and they take on none of the responsibility that came with producing it. They do not own the design, guarantee correctness, or maintain the system after it ships. The developer still does all three. The human effort in software engineering has therefore not disappeared. It has moved, and this paper traces where it went.

The popular name for the resulting practice is ``vibe coding'' \cite{Karpathy2025VibeCoding, Li2026VibeCoding}. A programmer manages an AI agent at a high level of intent and lets it produce the implementation. The term usually arrives with a promise of democratization, the idea that natural language can substitute for syntax and put software creation within reach of anyone who can describe what they want \cite{Ray2025VibeCodingReview, Sapkota2025VibeAgentic}. The practice has grown its own system. Rather than re-explain project conventions in every conversation, developers write persistent instruction files that an agent loads automatically, and public marketplaces already list more than 100,000 of these artifacts \cite{Hong2026SkillSmells}. Developers also pay per token, so an agent's work carries a metered cost that typing code never did.

Researchers have already mapped several pieces of what happens underneath the promise. Zamfirescu-Pereira et al. showed that natural-language specification is fragile even for sophisticated users, and later work documents how chronically under-specified ordinary task descriptions are \cite{Zamfirescu2023WhyJohnny, Pickering2025HowHumans}. Work on delegation and agency established that developers do not hand over control uniformly. They negotiate a boundary that shifts with task and expertise \cite{Khurana2025DoItForMe, Feng2026AgencyAI}. A newer line of work names the cost of drawing that boundary carelessly, measuring the erosion of recall and comprehension that follows when an AI produces work on a person's behalf \cite{Kosmyna2025CognitiveDebt, Bunch2026Trustworthiness}. Feng et al. come closest to our own delegation findings, establishing that agency allocation is stratified by expertise and task familiarity rather than tenure alone \cite{Feng2026AgencyAI}.

Each of these accounts was built where it could be built, in lab tasks and single-tool studies. What remains undocumented is how the pieces fit together across a working day. We do not know where the stratification Feng et al. describe comes from under field conditions and real cost pressure, or how developers author and maintain the instruction layer described above, now the default way to stabilize an otherwise fragile interaction \cite{Hong2026SkillSmells}. Nearly all of this evidence comes from high-resource contexts. Our study is set among professional developers in Bangladesh, a large and fast-growing software labor market that has received little research attention \cite{Sultana2026Bangladeshs} despite facing the same tools under tighter cost constraints than the populations usually studied.

We ran a three-phase mixed-methods study to address that gap. We surveyed 116 professional developers about their AI-assisted workflows. We then observed 14 developers build a realistic application with their own tools in contextual inquiry sessions, and interviewed each one immediately afterward. This design lets us triangulate what developers report against what they actually do.

Four research questions guide the study.

\begin{itemize}
    \item[] \textbf{RQ1:} When developers delegate programming tasks to AI coding assistants, where does the effort AI is supposed to eliminate reappear in their workflow?
    \item[] \textbf{RQ2:} Which tasks and decisions do developers consistently retain for themselves rather than delegate to AI, and what explains that boundary?
    \item[] \textbf{RQ3:} How do developers verify AI-generated code in practice, and what happens to their understanding and ownership of a system when verification is incomplete or bypassed?
    \item[] \textbf{RQ4:} What do these patterns imply for the design of AI-assisted development tools and for how developers are trained to use them?
\end{itemize}

\autoref{fig:vibecoding-teaser} summarizes the empirical story that emerged across these questions: code generation was readily delegated, but the work surrounding it did not disappear. Instead, effort shifted toward planning, instruction, cost management, and verification, while developers retained control over decisions they considered critical.

Effort did not disappear in any session we observed. It relocated in four consistent ways. Developers moved planning into files the agent reads automatically. They drew deliberate boundaries around work they refused to delegate. They treated token cost as an active constraint on what to accept and which model to run. And they let verification absorb most of the time generation had saved. That last displacement was the costliest one to get wrong. Nearly every respondent (99.1\%) had accepted AI-generated code that later proved to carry a subtle bug, and 93.1\% agreed an AI had solved a problem for them which they afterward could not explain. Careful reading was the single most connected behavior in the survey, associated with less automation bias and fewer accepted bugs, and comprehension of the resulting code tracked authorship or shared team convention rather than seniority.

We name the pattern underneath these results the competence paradox of vibe coding, and return to it in \S\ref{sec:discussion}. A tool marketed as flattening the expertise curve instead rewards a second, largely undocumented expertise in operating it well. That second expertise is most visible in the instruction files teams now write to encode their conventions and rarely validate.

This paper is findings-centric and makes three empirical contributions. First, a mixed-methods account of where AI-displaced effort lands in professional practice, and of the delegation boundaries developers draw and why. Second, evidence connecting incomplete verification to loss of code ownership and comprehension, together with shared team convention as an alternate route to comprehension that the cognitive debt literature has not considered. Third, an account of token cost as a first-class factor shaping code acceptance, model choice, and delegation rather than as background context.

For tool designers, our findings argue for making verification asymmetry visible in the interface, exposing cost at the moment of acceptance, and treating instruction files as software artifacts with the review and linting that implies. Researchers gain two variables current models leave out. Convention-based comprehension is a route to understanding that cognitive debt work does not include, and token cost belongs inside studies of acceptance rather than outside them. Teams and educators can read our data as evidence that AI-assisted development is a trainable competency which seniority does not reliably supply. We take the instruction-file implication furthest and build it into Better Skills, a proof-of-concept skill generator. It is an early design probe, not a system we evaluate here.

\section{Related Work} \label{sec:literature-review}
Large Language Models (LLMs) have changed how code is conceived, authored, and validated. We review that shift across seven threads, running from the emergence of vibe coding through delegation boundaries, prompting fragility, the instruction layer that grew up around it, the validation bottleneck, cognitive debt, and multi-agent systems.

\subsection{The Emergence of ``Vibe Coding'' and Displacement of Effort}

Andrej Karpathy coined the term ``vibe coding'' for a workflow that the democratization of LLMs has made ordinary, and the research literature has since formalized it \cite{Li2026VibeCoding, Karpathy2025VibeCoding, Horvat2025WhatIsVibe}. The programmer delegates the material production of code to an AI and manages the process at a high level of abstraction, ``giving in to the vibes'' \cite{Li2026VibeCoding, Ray2025VibeCodingReview}. The developer's role shifts from author of syntax to orchestrator of intent \cite{Drosos2025VibeCoding, Tholander2026VibeCoding}. Li et al. describe this as a ``material disengagement,'' in which the programmer operates through an abstract, probabilistic mediator rather than the code itself \cite{Li2026VibeCoding}.

Enthusiasts argue this lowers barriers to entry, and popular narratives frame it as the automation of programming outright \cite{Sapkota2025VibeAgentic}. The empirical record complicates that narrative. The ``vibe'' often amounts to surface plausibility, code that looks correct but carries subtle functional flaws \cite{Ray2025VibeCodingReview}. Observed in real sessions, developers do not surrender control but continuously monitor, steer, and repair the AI's output \cite{Drosos2025VibeCoding}. Behavioral modeling shows AI assistance displaces effort into reading, judging, and repairing rather than eliminating it \cite{Mozannar2024Reading}, and the time generation saves is substantially reabsorbed by verification and rework \cite{Ziegler2024Copilot, Vaithilingam2022Expectation}. Industry accounts name the same burden and call for AI that behaves less like a task-driven copilot and more like a symbiotic collaborator \cite{Hassan2026SE3}. Practical framings accordingly bound the paradigm to prototyping and low-stakes exploration \cite{Horvat2025WhatIsVibe, Voronin2024Development}.

As users distance themselves from execution logic they accumulate ``cognitive debt,'' a lack of understanding that makes future maintenance harder \cite{Li2026VibeCoding}. This worsens the ``paradox of the active user,'' where immediate output is prioritized over the learning needed to understand the tools in use \cite{Ray2025VibeCodingReview, Subramonyam2025Prototyping}. Feng et al. show agency shifting with task familiarity, with juniors over-delegating where their mental models are weakest \cite{Feng2026AgencyAI}. Interviews with UX professionals surface an almost identical rhythm of ideation, generation, debugging, and review, which suggests the phenomenon generalizes across roles \cite{Li2026VibeCoding}. This literature establishes that effort is displaced. Where it actually lands in professional development remains unmapped.

\subsection{Interaction Paradigms and Delegation Boundaries}
Barke et al. proposed a bimodal classification of programmer--LLM interaction. In \textit{acceleration} mode, the AI expedites tasks the developer already understands, working as a smart autocomplete judged against immediate expectation. In \textit{exploration} mode, the developer forages for unfamiliar APIs and patterns through back-and-forth dialogue \cite{Barke2023Grounded}.

Later work reframes this as agency allocation. ``Do It For Me'' paradigms maximize automation at the cost of comprehension and ownership, while ``Do It With Me'' paradigms preserve partnership at the cost of speed \cite{Khurana2025DoItForMe}. Full automation leaves users feeling like watchers of their own work, whereas confirmations, alternatives, and guided choices support partnership and co-adaptive learning \cite{Khurana2025DoItForMe, Gebreegziabher2025Supporting}. Users consistently seek an ``auteur'' role over end-to-end generation \cite{Earle2025DreamGarden, Voronin2024Development, Yen2025CodeShaping}, and autonomy preferences shift continuously with task, context, and confidence \cite{Wong2025ItIsASpectrum}. Expertise stratifies the boundary. Seniors maintain control through small, testable delegation and minimal diffs, while juniors oscillate between over-reliance and defensive resistance \cite{Feng2026AgencyAI}, and novice-facing tools must scaffold competence rather than automate the struggle away \cite{Ali2025EvenThough, Zheng2025NonAI, Mowar2025CodeA11y}. Proactive assistants complicate the boundary further, arriving as disruption when they cross a line the developer did not draw \cite{Chen2025NeedHelp, Pu2025Assistance}, and the role assigned to the AI reshapes how its contributions are valued \cite{Schecter2025Role}.

What this literature leaves open is how developers operationalize the delegation boundary in unconstrained settings, and which activities they refuse to delegate. Our contextual inquiries take up that question in the field.

\subsection{Prompt Engineering as a Fragile Interface}
As natural language becomes the primary programming interface, the literature converges on a single finding. Prompting is fragile even for sophisticated users. Non-AI experts cannot envision how the model will interpret their words, a problem named the ``Gulf of Envisioning'' \cite{Geng2025Exploring, Zamfirescu2023WhyJohnny, Subramonyam2024Bridging}. Natural task descriptions are chronically under-specified \cite{Pickering2025HowHumans}, output quality tracks articulation quality rather than task difficulty \cite{Wang2024UML}, and even incidental context such as TODO comments measurably alters output \cite{OBrien2024Todo}.

The difficulties take several forms. Users face a cold-start ``blank page'' paralysis and benefit from structured scaffolding such as pre-defined design questions \cite{Subramonyam2025Prototyping, Zamfirescu2025Beyond}. They struggle to balance generality and specificity, since vague prompts yield generic boilerplate while overly specific ones constrain or overfit \cite{Subramonyam2025Prototyping, Pickering2025HowHumans}. And without a gold standard, they iterate blindly, ``prompting in the dark'' while losing track of what worked \cite{He2025Prompting, Subramonyam2025Prototyping}.

Responses include visual environments for chaining prompts, conversation regression testing, and interfaces separating design questions from confirmed requirements \cite{Zamfirescu2025Beyond, Epperson2025Interactive, Wong2025ItIsASpectrum}, along with prompts treated as shared, versionable team artifacts \cite{Subramonyam2025Prototyping}. Together, this constitutes context engineering, a skill package the literature has anticipated \cite{Subramonyam2025Prototyping, Zamfirescu2025Beyond} but not yet documented in the field.

\subsection{Persistent Instruction and the Question of Skill Quality}
The most concrete realization of this anticipated skill package is appearing in practice, if not yet in the literature. Rather than re-explaining conventions in every conversation, developers increasingly package procedural knowledge and project-specific context into portable files an agent loads automatically. These include CLAUDE.md, AGENTS.md, and the Agent Skill, a directory built around a required SKILL.md file \cite{Hong2026SkillSmells}. Work on agent-computer interfaces argues that how an agent is given commands, files, and structured context is a first-class design problem in its own right \cite{Yang2024SWEagent}. Work on reducing long agent trajectories \cite{Xiao2026TrajectoryReduction} and simplifying code passed to models \cite{Zhang2022DietCode} raises a parallel concern. What an agent holds in context shapes both cost and reliability.

Despite this adoption, almost nothing is known about whether these files are any good. The first systematic study of the artifact analyzed 238 real-world SKILL.md files and found authoring guidance scattered and incomplete. A multivocal review of 29 sources yielded only 26 best practices covering seven of thirteen content categories \cite{Hong2026SkillSmells}. Inverting these practices into \textit{skill smells}, by analogy to code smells, the authors found over 99\% of files contained at least one smell, averaging 10.5 each, and smells introduced early rarely got corrected \cite{Hong2026SkillSmells}. The files developers rely on to keep an agent aligned with their conventions are largely unvalidated and, once flawed, tend to stay that way.

Prompt engineering has drawn sustained HCI attention, and code quality has decades of research behind it. This instruction layer has almost neither. How teams decide what belongs in a shared skill, how it is maintained, and what separates a skill that helps from one that misleads remain open questions we take up directly.

\subsection{The Validation Bottleneck: Trust, ``Vibes,'' and Verification}
The most critical challenge is determining whether generated code actually works. The vibe coding mindset encourages reliance on surface plausibility, and code that looks right is often accepted without rigorous review \cite{Ray2025VibeCodingReview}. Automation bias supports this behavior, since confidently presented output is trusted over one's own judgment \cite{Goddard2012AutomationBias, Li2025From}. Juniors accept outputs that pass superficial checks, while seniors emphasize minimal diffs, test-bounded changes, and manual verification \cite{Feng2026AgencyAI}.

Research identifies two primary validation strategies \cite{Ferdowsi2024Validating}. Execution is effective but infeasible for partial snippets or scaffolding. Examination means reading and mentally simulating code, which AI-generated code invites developers to skim rather than do, letting subtle logic errors and vulnerabilities pass \cite{Pickering2025HowHumans, Yan2024Ivie}. Tools like Ivie anchor in-place explanations to help users verify intent \cite{Yan2024Ivie}, and live programming environments surface runtime values to reduce the load of mental simulation \cite{Ferdowsi2024Validating}. But validation is cognitive as much as technical. Absent ground truth, users convince themselves the output is correct because they cannot prove otherwise \cite{He2025Prompting}. Code that looks right is accepted, with defects surfacing later \cite{Feng2026AgencyAI, Vaithilingam2022Expectation}, and acceptance feels like progress \cite{Ziegler2024Copilot} even when verification absorbs the gains \cite{Mozannar2024Reading}.

What remains unexamined is the moment of acceptance itself. What happens when time pressure, token costs, or weak understanding lead developers to accept unverified code, and how do they recover?

\subsection{Cognitive Debt and the Erosion of Understanding}
The downstream cost of weak validation has acquired a name. Cognitive debt is the measurable erosion of recall, comprehension, and ownership that accumulates when an AI produces work on a person's behalf \cite{Kosmyna2025CognitiveDebt, Bunch2026Trustworthiness}. Demonstrated experimentally in AI-assisted writing \cite{Kosmyna2025CognitiveDebt}, it converges with warnings that material disengagement degrades capacity for maintenance and debugging \cite{Li2026VibeCoding, Ray2025VibeCodingReview}. A parallel strand documents the same erosion at the codebase level, where AI-generated code introduces technical debt across most established categories, often invisible until someone other than the prompter must extend the system \cite{Ehsani2026TechnicalDebt}. The educational parallel is well documented, since learners who delegate the struggle complete everything while learning little \cite{Ali2025EvenThough}, and non-experts lack the grounding to judge AI contributions at all \cite{Zheng2025NonAI, Mowar2025CodeA11y}.

Co-adaptive approaches, in which human and AI refine shared understanding \cite{Gebreegziabher2025Supporting}, and organizational mechanisms such as prompt-and-code reviews \cite{Feng2026AgencyAI} offer countervailing directions. But no study has traced which practices prevent cognitive debt outside the classroom. Our study measures it through task success, error recovery, and developers' ability to modify or explain generated code, stratified by seniority. Where learning resources are limited, as in the Global South, preventing it is essential to building local expertise.

\subsection{Beyond Chat: Multi-Agent Systems and Future Interfaces}
The limits of single-turn chat are driving multi-agent architectures. A planner agent decomposes a high-level prompt into sub-tasks executed by specialized coder, designer, or debugger agents, which shifts debugging from code syntax to inter-agent conversation \cite{Epperson2025Interactive, Earle2025DreamGarden}. Purpose-built agent-computer interfaces substantially change how reliably an agent completes a task, independent of the underlying model \cite{Yang2024SWEagent}.

This work stays largely at the level of architectures rather than asking how developers interact with such systems under real constraints. Who uses multi-model workflows, and how do token costs drive tool switching? Tracing real tool use across a diverse population, including the Global South, can show whether current ecosystems serve developers equitably.

\subsection{Summary and Research Gap}
Across these threads, a single pattern emerges. AI assistance does not remove human effort from software engineering; it relocates it. Effort moves into monitoring and repair \cite{Drosos2025VibeCoding, Mozannar2024Reading, Vaithilingam2022Expectation}, into expertise-stratified delegation \cite{Barke2023Grounded, Khurana2025DoItForMe, Wong2025ItIsASpectrum, Feng2026AgencyAI}, into context artifacts whose own quality has barely been studied \cite{Subramonyam2025Prototyping, He2025Prompting, Zamfirescu2023WhyJohnny, Hong2026SkillSmells}, and into a validation bottleneck where trust and bias operate largely unchecked \cite{Li2025From, Ferdowsi2024Validating, Yan2024Ivie}. When relocation fails, it turns into cognitive debt at both the individual and codebase level \cite{Ali2025EvenThough, Kosmyna2025CognitiveDebt, Ehsani2026TechnicalDebt}.

This evidence base rests largely on controlled lab tasks, single tools, and student or novice populations \cite{Barke2023Grounded, Geng2025Exploring, Ali2025EvenThough, Vaithilingam2022Expectation}. Field studies of professional engineers using their own stacks, with observed behavior triangulated against self-report at scale, remain scarce. The practices our data surface are fragmentarily documented or absent, including retained scaffolding, plan-first staging, uneven-quality context artifacts, and token-driven acceptance. Meanwhile, vision papers call for AI collaborators that understand project-specific intent \cite{Hassan2026SE3}, and the one empirical study of the closest existing artifact finds it riddled with quality problems \cite{Hong2026SkillSmells}.

Our mixed-methods study addresses this gap directly. RQ1 and RQ2 trace where displaced effort and delegation boundaries actually land in professional practice. RQ3 follows validation failures through to cognitive debt and recovery. RQ4 turns these field patterns into concrete guidance for tool design and training.

\section{Methodology} \label{sec:methodology}
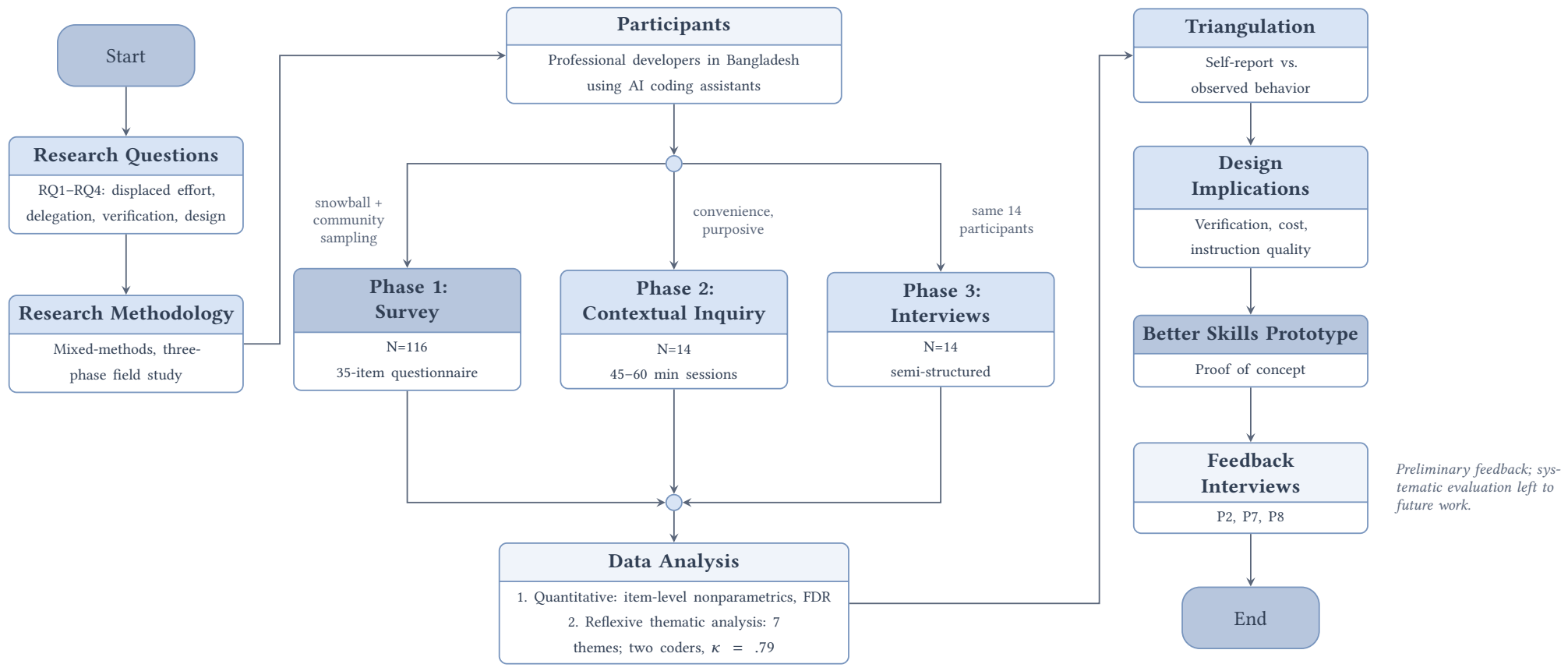
\begin{figure*}[t]
    \centering
    \resizebox{0.92\textwidth}{!}{%

\colorlet{softblue}{paperPale}
\colorlet{softgreen}{paperLight}
\colorlet{softlavender}{paperPale}
\colorlet{softcream}{paperWash}
\colorlet{softrose}{paperLight}
\colorlet{softcyan}{paperPale}

\colorlet{softborder}{paperMid}
\colorlet{softtext}{paperInk}
\colorlet{softsecondary}{paperMuted}

\begin{tikzpicture}[
    >=stealth,
    line width=0.55pt,
    spinebox/.style={
        rectangle split,
        rectangle split parts=2,
        rectangle split part fill={softblue,white},
        draw=softborder,
        rounded corners=3pt,
        align=center,
        inner sep=3.5pt,
        font=\small,
        text=softtext,
        text width=3.0cm
    },
    databox/.style={
        rectangle split,
        rectangle split parts=2,
        rectangle split part fill={softcream,white},
        draw=softborder,
        rounded corners=3pt,
        align=center,
        inner sep=3.5pt,
        font=\small,
        text=softtext,
        text width=2.9cm
    },
    surveybox/.style={
        databox,
        rectangle split part fill={softgreen,white}
    },
    inquirybox/.style={
        databox,
        rectangle split part fill={softblue,white}
    },
    interviewbox/.style={
        databox,
        rectangle split part fill={softlavender,white}
    },
    analysisbox/.style={
        databox,
        rectangle split part fill={softcream,white}
    },
    tribox/.style={
        spinebox,
        rectangle split part fill={softlavender,white}
    },
    implicationbox/.style={
        spinebox,
        rectangle split part fill={softblue,white}
    },
    prototypebox/.style={
        spinebox,
        rectangle split part fill={softgreen,white}
    },
    feedbackbox/.style={
        spinebox,
        rectangle split part fill={softcream,white}
    },
    startstop/.style={
        draw=softborder,
        rounded corners=7pt,
        align=center,
        font=\small,
        text=softtext,
        minimum width=1.9cm,
        minimum height=0.85cm
    },
    mdot/.style={
        circle,
        draw=softborder,
        fill=softcyan,
        inner sep=2.2pt
    },
    arr/.style={
        ->,
        draw=softsecondary
    },
    samplab/.style={
        font=\scriptsize,
        align=center,
        text=softsecondary
    },
    sidenote/.style={
        font=\scriptsize\itshape,
        align=left,
        text=softsecondary,
        text width=2.5cm
    }
]


\node[startstop, fill=softgreen] (start) at (0,7.8)
    {Start};

\node[spinebox] (rq) at (0,6.0)
    {\textbf{Research Questions}
    \nodepart{two}
    \scriptsize RQ1--RQ4: displaced effort, delegation, verification, design};

\node[spinebox] (design) at (0,3.8)
    {\textbf{Research Methodology}
    \nodepart{two}
    \scriptsize Mixed-methods, three-phase field study};


\node[databox, text width=4.4cm] (part) at (7.6,7.8)
    {\textbf{Participants}
    \nodepart{two}
    \scriptsize Professional developers in Bangladesh using AI coding assistants};

\node[mdot] (dot1) at (7.6,6.3) {};


\node[surveybox] (s1) at (3.9,4.0)
    {\textbf{Phase 1:\\Survey}
    \nodepart{two}
    \scriptsize N=116\\
    35-item questionnaire};

\node[inquirybox] (s2) at (7.6,4.0)
    {\textbf{Phase 2:\\Contextual Inquiry}
    \nodepart{two}
    \scriptsize N=14\\
    45--60 min sessions};

\node[interviewbox] (s3) at (11.3,4.0)
    {\textbf{Phase 3:\\Interviews}
    \nodepart{two}
    \scriptsize N=14\\
    semi-structured};


\node[mdot] (dot2) at (7.6,1.6) {};

\node[analysisbox, text width=4.6cm] (analysis) at (7.6,0.2)
    {\textbf{Data Analysis}
    \nodepart{two}
    \scriptsize
    1. Quantitative: item-level nonparametrics, FDR\\
    2. Reflexive thematic analysis: 7 themes; two coders, $\kappa=.79$};


\node[tribox] (tri) at (15.6,7.8)
    {\textbf{Triangulation}
    \nodepart{two}
    \scriptsize Self-report vs.\\
    observed behavior};

\node[implicationbox] (imp) at (15.6,5.7)
    {\textbf{Design\\Implications}
    \nodepart{two}
    \scriptsize Verification, cost,\\
    instruction quality};

\node[prototypebox] (proto) at (15.6,3.7)
    {\textbf{Better Skills Prototype}
    \nodepart{two}
    \scriptsize Proof of concept};

\node[feedbackbox] (fb) at (15.6,1.8)
    {\textbf{Feedback\\Interviews}
    \nodepart{two}
    \scriptsize P2, P7, P8};

\node[startstop, fill=softrose] (end) at (15.6,0.0)
    {End};


\node[sidenote, anchor=west] at (17.5,1.8)
    {Preliminary feedback; systematic evaluation left to future work.};


\draw[arr] (start) -- (rq);
\draw[arr] (rq) -- (design);

\draw[arr]
    (design.east)
    -- ++(0.5,0)
    |- (part.west);

\draw[arr] (part.south) -- (dot1);

\draw[arr] (dot1) -| (s1.north);
\draw[arr] (dot1) -- (s2.north);
\draw[arr] (dot1) -| (s3.north);


\node[samplab, anchor=east] at (3.7,5.5)
    {snowball +\\community\\sampling};

\node[samplab, anchor=west] at (7.75,5.5)
    {convenience,\\purposive};

\node[samplab, anchor=west] at (11.45,5.5)
    {same 14\\participants};


\draw[arr] (s1.south) |- (dot2);
\draw[arr] (s2.south) -- (dot2);
\draw[arr] (s3.south) |- (dot2);

\draw[arr] (dot2) -- (analysis.north);


\draw[arr]
    (analysis.east)
    -- (13.5,0.2)
    -- (13.5,7.8)
    -- (tri.west);


\draw[arr] (tri) -- (imp);
\draw[arr] (imp) -- (proto);
\draw[arr] (proto) -- (fb);
\draw[arr] (fb) -- (end);

\end{tikzpicture}
    }

    \caption{Study design and analytical workflow. Three data-collection phases feed quantitative and qualitative analysis, triangulation, design implications, the Better Skills proof of concept, and feedback interviews.}

    \Description{A left-to-right and top-to-bottom flow diagram. Start leads to Research Questions and Research Methodology on the left. A Participants box at the top center identifies professional developers in Bangladesh using AI coding assistants. It branches into three side-by-side boxes: Phase 1 Survey, N=116 and 35 items; Phase 2 Contextual Inquiry, N=14 and 45–60-minute sessions; and Phase 3 Interviews, N=14 and semi-structured. The three phases converge on Data Analysis, which lists quantitative nonparametric analysis with false-discovery-rate correction and reflexive thematic analysis with seven themes, two coders, and kappa of .79. An arrow then moves to a right-hand vertical sequence of Triangulation, Design Implications, Better Skills Prototype, Feedback Interviews with P2, P7, and P8, and End.}

    \label{fig:process-diagram}
\end{figure*}

We conducted a three-phase, mixed-methods study to answer our research questions. It comprised a large-scale survey (N=116), contextual inquiry sessions (N=14), and semi-structured post-task interviews. We then developed a prototype and gathered feedback from three returning participants. \autoref{fig:process-diagram} summarizes the process.

\subsection{Participants and Recruitment}
\label{sec:participants}

\begin{table*}[t]
    \centering
    \caption{Demographic composition of survey respondents (N=116), limited to the variables used elsewhere in this chapter's statistical analysis.}
    \label{tab:sample-demographics}
    \small
    \begin{tabular}{@{}llcc@{}}
        \toprule
        \textbf{Category} & \textbf{Group} & \textbf{n} & \textbf{\%} \\
        \midrule
        \multirow{6}{*}{\parbox{2.6cm}{Role}}
            & Software engineer / developer & 89 & 76.7 \\
            & Student (CS or related field) & 13 & 11.2 \\
            & Teacher / Educator / Faculty Member & 4 & 3.4 \\
            & Product / Project manager & 3 & 2.6 \\
            & Data scientist / ML engineer & 3 & 2.6 \\
            & Other (Computer Vision Engineer, Researcher, QA, Designer; $n=1$ each) & 4 & 3.4 \\
        \midrule
        \multirow{6}{*}{\parbox{2.6cm}{Years of programming experience}}
            & Less than 1 year & 10 & 8.6 \\
            & 1--3 years & 31 & 26.7 \\
            & 4--7 years & 31 & 26.7 \\
            & 8--15 years & 26 & 22.4 \\
            & More than 15 years & 15 & 12.9 \\
            & Not applicable & 3 & 2.6 \\
        \midrule
        \multirow{4}{*}{\parbox{2.6cm}{Coding proficiency}}
            & Advanced (designs systems, reviews code, mentors others) & 52 & 44.8 \\
            & Intermediate (builds functional applications independently) & 44 & 37.9 \\
            & Beginner (writes simple scripts with guidance) & 19 & 16.4 \\
            & Non-coder & 1 & 0.9 \\
        \midrule
        \multirow{5}{*}{\parbox{2.6cm}{Tenure using AI coding assistants}}
            & Less than 3 months & 10 & 8.6 \\
            & 3--6 months & 13 & 11.2 \\
            & 6--12 months & 35 & 30.2 \\
            & 1--2 years & 42 & 36.2 \\
            & More than 2 years & 16 & 13.8 \\
        \midrule
        \multirow{4}{*}{\parbox{2.6cm}{Usage frequency on a typical working day}}
            & A few times per week & 6 & 5.2 \\
            & Once or twice a day & 17 & 14.7 \\
            & A few times per day & 43 & 37.1 \\
            & Multiple times per hour & 50 & 43.1 \\
        \bottomrule
    \end{tabular}
\end{table*}

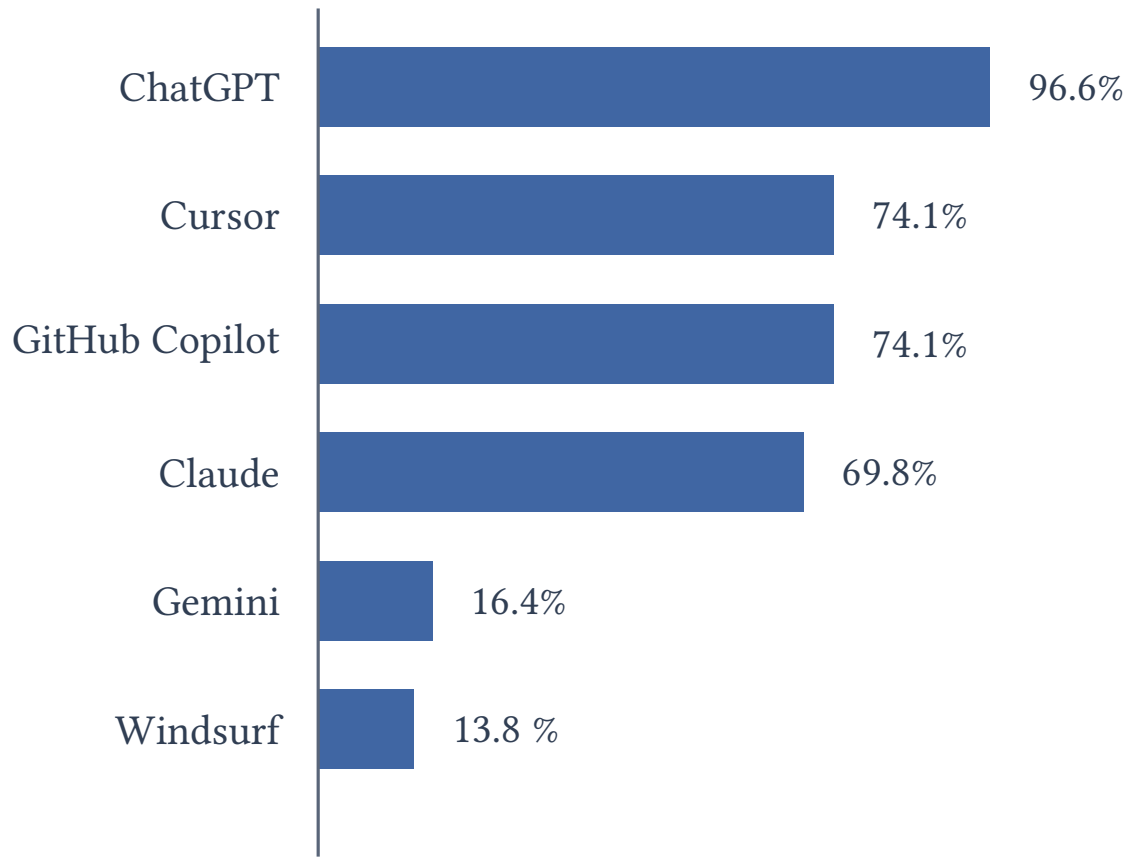
\begin{figure}[t]
    \centering
    \begin{tikzpicture}[x=0.046cm, y=-0.85cm]
        \colorlet{tooladopt}{paperBlue}

        \foreach \barlabel/\val [count=\i from 0] in {
            ChatGPT/96.6,
            Cursor/74.1,
            {GitHub Copilot}/74.1,
            Claude/69.8,
            Gemini/16.4,
            Windsurf/13.8
        } {
            \fill[tooladopt] (0,\i) rectangle (\val,\i+0.62);
            \node[anchor=east, font=\small, xshift=-4pt] at (0,\i+0.31) {\barlabel};
            \node[anchor=west, font=\small, xshift=4pt] at (\val,\i+0.31) {\val\%};
        }

        \draw[paperMuted,line width=0.6pt] (0,-0.3) -- (0,6.3);
    \end{tikzpicture}

    \caption{Adoption of AI coding assistants among survey respondents (N=116) over the prior three months. Respondents could select more than one tool; three additional tools (a Duck.ai-hosted model, DeepSeek, and Antigravity) were each named by a single respondent (0.9\%) and are omitted from the chart.}

    \Description{Adoption of AI coding assistants among survey respondents (N=116) over the prior three months. Respondents could select more than one tool; three additional tools (a Duck.ai-hosted model, DeepSeek, and Antigravity) were each named by a single respondent (0.9\%) and are omitted from the chart.}

    \label{fig:tool-adoption}
\end{figure}

We recruited software engineers and developers in Bangladesh who actively use at least one AI coding assistant in their regular work, such as ChatGPT, Claude, GitHub Copilot, Cursor, or Windsurf \cite{openai_chatgpt, anthropic_claude, github_copilot, cursor, windsurf}. The survey was seeded through professional networks, with respondents encouraged to forward it, and shared in public Facebook communities for Bangladeshi software engineers. Survey respondents (N=116) represented frontend, backend, full-stack, and adjacent engineering roles across industries and experience levels. Contextual inquiry participants were recruited separately through convenience sampling from our professional network. \autoref{tab:sample-demographics} summarizes the variables used later in the analysis, namely role, programming experience, coding proficiency, AI-assistant tenure, and daily usage frequency. Most respondents were software engineers (76.7\%), with at least one year of programming experience (88.8\%). Most also reported intermediate or advanced coding skills (82.8\%). AI-assistant experience was shorter, and half reported one year or less. However, most used AI assistants several times daily (80.2\%). ChatGPT was most widely adopted (96.6\%), followed by Cursor and GitHub Copilot (74.1\% each), Claude (69.8\%), Gemini (16.4\%), and Windsurf (13.8\%). Respondents could select multiple tools (\autoref{fig:tool-adoption}).

We conducted 11 sessions, analyzing data iteratively during collection. After no substantially new themes emerged, we added three more (N=14) to confirm saturation and stability \cite{Saunders2018Saturation}. Each participant completed a semi-structured post-task interview immediately after their session. From the convenience pool, we purposively selected participants \cite{Patton1990Qualitative} to vary in seniority and primary tool choice, which reflected chat-first versus agent-first workflows. Saturation, not a predetermined sample size, governed recruitment stopping \cite{Saunders2018Saturation}.

We stratified participants into three experience levels. \textit{Junior} participants were recent graduates, interns, or developers with under one year of experience. \textit{Mid-level} participants had one to three years of industry experience. \textit{Senior} participants had 4+ years of experience and routinely made architectural decisions for products or teams. Two senior software engineers independently reviewed these assignments for consistency with Bangladeshi hiring practices. Disagreements were resolved through discussion. \autoref{tab:contextual_inquiry_participants} summarizes participant distribution across levels.

\begin{table*}[t]
    \centering
    \caption{Participant characteristics in the contextual inquiry study.}
    \label{tab:contextual_inquiry_participants}
    \small
    \begin{tabular}{lll}
        \hline
        \textbf{Participant} & \textbf{Experience Level} & \textbf{Background and Relevant Experience} \\
        \hline
        P1  & Junior & Fresh graduate; bootcamp-trained; real-world projects \\
        P2  & Junior & Final year university student; self-learner; user-facing projects \\
        P3  & Junior & Software engineer inter; bootcamp-trained; user-facing projects \\
        P4  & Mid & Software engineer; 2 years; self-learner; user-facing projects \\
        P5  & Mid & Product Engineer; 2 years; self-learning developer; user-facing projects \\
        P6  & Junior & Frontend developer; fresh graduate; self-learner \\
        P7  & Senior & Frontend-focused software engineer; 14+ years; project lead \\
        P8  & Mid & Software engineer; 2 years; creates custom AI skills \\
        P9  & Senior & Software engineer; 6+ years; backend-focused \\
        P10 & Senior & Software engineer; 6 years; full-stack; architectural decisions \\
        P11 & Mid & Software engineer; 2 years; full-stack \\
        P12 & Senior & Software engineer; 2 years; full-stack \\
        P13 & Junior & Software engineer; 6 months; frontend-focused; real-world projects \\
        P14 & Mid & Software engineer; 2 years; backend-focused \\
        \hline
    \end{tabular}
    
    \vspace{2pt}
    \parbox{\textwidth}{\footnotesize
    \textit{Note:} All participants reported using agentic AI tools during software development, with variation in their level and patterns of use.
    }
\end{table*}

\subsection{Phase 1: Survey}
\label{sec:survey}

The online survey took approximately 15 minutes and contained 35 questions, including multiple-choice, ranking, Likert-scale, and open-ended items \cite{likert1932technique, mackenzie2013human}. The instrument covered eight areas: (1) background, role, experience, proficiency, and tools; (2) usage patterns, frequency, delegated tasks, and chat-versus-agentic use; (3) prompting, planning, decomposition, refinement, template reuse, and difficulties such as blank-page paralysis; (4) verification and trust, including reading habits, strategies, automation bias, and subtle defects; (5) self-efficacy, agency, workflow, perceived skill loss, ownership, and cognitive debt; (6) perceived productivity and code quality; (7) failure modes and desired improvements, including open-ended accounts; and (8) cost, payment responsibility, monthly spending, and perceived justification.

Item construction drew on prior work on programmer--LLM interaction, including acceleration and exploration modes \cite{Barke2023Grounded}, prompting in the dark \cite{He2025Prompting}, automation bias \cite{Li2025From}, agency allocation between ``Do It For Me'' and ``Do It With Me'' \cite{Khurana2025DoItForMe, Wong2025ItIsASpectrum}, and cognitive debt \cite{Kosmyna2025CognitiveDebt}. We piloted the instrument with developers to refine wording, ordering, and length before launch.

The 40 Likert items included 14 reverse-coded items and were grouped into five thematic blocks, namely Prompting Style, Trust \& Automation Bias, Self-Efficacy \& Agency, Workflow Integration, and Productivity \& Quality. Quantitative results in \S\ref{sec:quant-findings} are reported at the item level (\S\ref{sec:analysis}), and block labels indicate questionnaire coverage.

\subsection{Phase 2: Contextual Inquiry}
\label{sec:contextual-inquiry}

Self-reported verification and delegation can be unreliable. The second phase therefore observed developers rather than asking how they work. We conducted 45--60-minute contextual inquiry sessions across experience levels, remotely via Google Meet. Participants shared and recorded their full screens with consent. Each participant built a small but complete management-system web application with realistic requirements. They used their own technology stack and AI tools, such as Claude with VS Code or ChatGPT and Claude via chat interfaces. The task included frontend and backend variants, and participants could complete both as full-stack. To minimize environment setup, participants used \texttt{localStorage} for frontend data or file-based databases such as JSON or SQLite for backend data. No other constraints were imposed.

Following contextual inquiry practice \cite{Beyer1998Contextual}, participants thought aloud while working \cite{Ericsson1984Protocol}, and the interviewer asked questions when needed. Two authors conducted each session, one leading and probing decisions, the other taking notes. The session was documented immediately afterward while details remained fresh. The observation protocol covered seven dimensions, namely task understanding and planning, prompting, AI-output integration, failure response, verification, architectural decisions, and tool switching. Midway through each session, participants explained, fixed, or changed a concrete part of their application, framed as a natural follow-up. This assessed whether they understood and owned the code rather than merely generating it. For each session, we collected complete prompt logs, the Git repository and commit history, and screen recordings. The observer also rated each session using a skills rubric covering planning, prompting, verification, debugging, tool selection, architectural decision-making, and adaptability, enabling structured comparison across experience levels.

\subsection{Phase 3: Post-Task Interviews}
\label{sec:interviews}

Immediately after each session, the same two authors conducted a semi-structured interview. It covered typical AI use, delegation preferences, difficulties, trust and verification decisions, and when participants proceeded independently. It also asked what they would change about their AI-assisted workflow and what distinguishes effective AI-assisted development. Interviews were recorded and transcribed. Transcripts were analyzed alongside corresponding session notes and artifacts.

\subsection{Data Analysis}
\label{sec:analysis}

\subsubsection{Quantitative Analysis}

For survey data, we computed descriptive statistics for usage frequency, task delegation, tool adoption, verification strategies, and failure modes, comparing experience-level subgroups. Inferential analysis used item-level nonparametric tests. We ran Spearman rank correlations between ordinal items and Kruskal-Wallis tests across experience bands, with Benjamini-Hochberg FDR correction for each reported test family. We treat these as exploratory associations in a cross-sectional self-report sample and draw no causal conclusions.

\subsubsection{Qualitative Analysis}

We analyzed open-ended survey responses, interview transcripts, and observation notes using reflexive thematic analysis (RTA) \cite{Braun2019Reflexive} and affinity diagramming \cite{Courage2005Field}. Open-ended survey items were optional. 27 respondents answered at least one, and these responses were analyzed alongside interview and observation data. Theme development followed RTA's interpretive process. Both authors independently open-coded an initial subset, then developed candidate themes through discussion rather than a fixed codebook. This follows Braun and Clarke's view that analyst judgment is part of the instrument, not an error to measure \cite{Braun2019Reflexive}. After finalizing the themes, we conducted a separate consistency check. Both authors independently assigned 359 claims to one of seven themes, yielding strong agreement (Cohen's $\kappa=0.79$) \cite{Cohen1960Kappa}. Disagreements were resolved through negotiated agreement \cite{Campbell2013Coding}. We report $\kappa$ only as a labeling consistency check, not as evidence of RTA validity, which depends on interpretive coherence and depth. Finally, we organized codes on an affinity diagramming board, grouping related codes into sub-themes and those into higher-level themes.

\subsubsection{Triangulation}
\label{sec:triangulation}

Finally, we compared the three strands following mixed-methods integration practice \cite{Fetters2013Integration, OCathain2010Three}. When the survey indicated code verification, we examined whether it occurred during sessions. Conversely, when sessions revealed behaviors such as accepting unread code, we returned to the survey to estimate prevalence. Each check was classified as convergent, complementary, tension, or silent, where silent means no corresponding item existed in the other strand. We report all four as substantive results rather than treating agreement alone as informative (\S\ref{sec:findings-triangulation}). When self-reports diverged from observed behavior, we treated the mismatch as an analytic finding rather than measurement error, because this gap is directly relevant to tool design and training (RQ4).

\subsection{Prototype Development}
\label{sec:prototype-development}

Based on the three study phases, we developed Better Skills, a prototype addressing the instruction-file quality gap they surface, described alongside the design implications it instantiates (\S\ref{sec:better-skills}). For preliminary feedback, we walked three contextual inquiry participants (P2, P7, P8) through the prototype in short follow-up interviews, asking each to configure and inspect a generated skill.

\subsection{Ethical Considerations}
\label{sec:ethics}

The study was approved by our institutional review board. We obtained verbal consent from every participant before their session, and participants could withdraw at any time. Survey responses were collected anonymously, and respondents were asked for contact details only if interested in follow-up. Session and interview data were pseudonymized before analysis. Participants shared only non-proprietary prompts and code, and identifying material was excluded from the dataset.

\section{Findings} \label{sec:findings}
We first summarize the quantitative patterns, then the qualitative themes, and close by triangulating the two strands. Quotes are attributed by ID, with P1--P14 for contextual inquiry participants and SP1--SP27 for survey respondents.

\subsection{Quantitative Findings}
\label{sec:quant-findings}

We report results at the item level with nonparametric association tests (Spearman's $\rho$; Kruskal-Wallis) and Benjamini-Hochberg FDR correction across the reported family (\S\ref{sec:analysis}). These are exploratory associations in a cross-sectional self-report sample, not causal estimates.

\subsubsection{Prompting Is Disciplined, but Feedback Is Missing}

Structured prompting habits were widespread. 60.3\% usually or almost always planned a prompt before writing it, and 82.8\% often decomposed complex tasks into smaller sequential prompts. What respondents lacked was not discipline but feedback on whether it was working. 88.8\% reported at least occasionally ``prompting in the dark,'' varying wording without knowing what helped, and 26.7\% experienced this frequently. Half reported blank-page paralysis a few times monthly.

\subsubsection{Verification Is Uneven and Tracks Every Bad Outcome We Measured}

\begin{figure}[t]
    \centering
    \begin{tikzpicture}[x=0.038cm, y=-0.72cm]
        \colorlet{verifbar}{paperBlue}

        \node[font=\small\bfseries, anchor=east] at (0,-1) {Validation strategies used};

        \foreach \barlabel/\val [count=\i from 0] in {
            {Run it, check output (execution)}/73.3,
            {Mentally trace logic (examination)}/66.4,
            {Compare with documentation}/65.5,
            {Ask AI to explain the code}/39.7,
            {Unit tests}/26.7,
            {Peer review}/20.7,
            {No systematic verification}/6.9
        } {
            \fill[verifbar] (0,\i) rectangle (\val,\i+0.58);
            \node[anchor=east, font=\footnotesize, xshift=-3pt] at (0,\i+0.29) {\barlabel};
            \node[anchor=west, font=\footnotesize, xshift=3pt] at (\val,\i+0.29) {\val\%};
        }
        \draw[paperMuted,line width=0.6pt] (0,-0.3) -- (0,7.0);
    \end{tikzpicture}
    \hfill
    \begin{tikzpicture}[x=0.038cm, y=-0.72cm]
        \colorlet{failbar}{paperMid}

        \node[font=\small\bfseries, anchor=east] at (0,-1) {Failure modes personally experienced};

        \foreach \barlabel/\val [count=\i from 0] in {
            {Confidently outdated/deprecated answer}/75.9,
            {Syntactically correct, logically wrong}/70.7,
            {Debugging cost exceeded time saved}/47.4,
            {Misunderstood intent, wrong problem}/37.9,
            {Iterative refinements made it worse}/26.7,
            {Could not understand code well enough to debug}/20.7,
            {Different answers to the same prompt}/14.7,
            {Accepted a security vulnerability}/9.5
        } {
            \fill[failbar] (0,\i) rectangle (\val,\i+0.58);
            \node[anchor=east, font=\footnotesize, xshift=-3pt] at (0,\i+0.29) {\barlabel};
            \node[anchor=west, font=\footnotesize, xshift=3pt] at (\val,\i+0.29) {\val\%};
        }
        \draw[paperMuted,line width=0.6pt] (0,-0.3) -- (0,8.0);
    \end{tikzpicture}

    \caption{Self-reported validation strategies (left) and personally experienced failure modes of AI-generated code (right) among survey respondents (N=116). Both items were multi-select; percentages are of all respondents.}

    \Description{Two horizontal bar charts. The left chart shows validation strategies: execution-based checking is most common at 73.3 percent, followed by mental tracing at 66.4 percent and documentation comparison at 65.5 percent; unit tests and peer review are used by roughly a quarter and a fifth of respondents, and 6.9 percent report no systematic verification. The right chart shows failure modes: 75.9 percent experienced confidently outdated answers, 70.7 percent logically wrong but syntactically correct code, 47.4 percent spent more time debugging than the AI saved, and 9.5 percent accepted a security vulnerability.}

    \label{fig:verification-failures}
\end{figure}
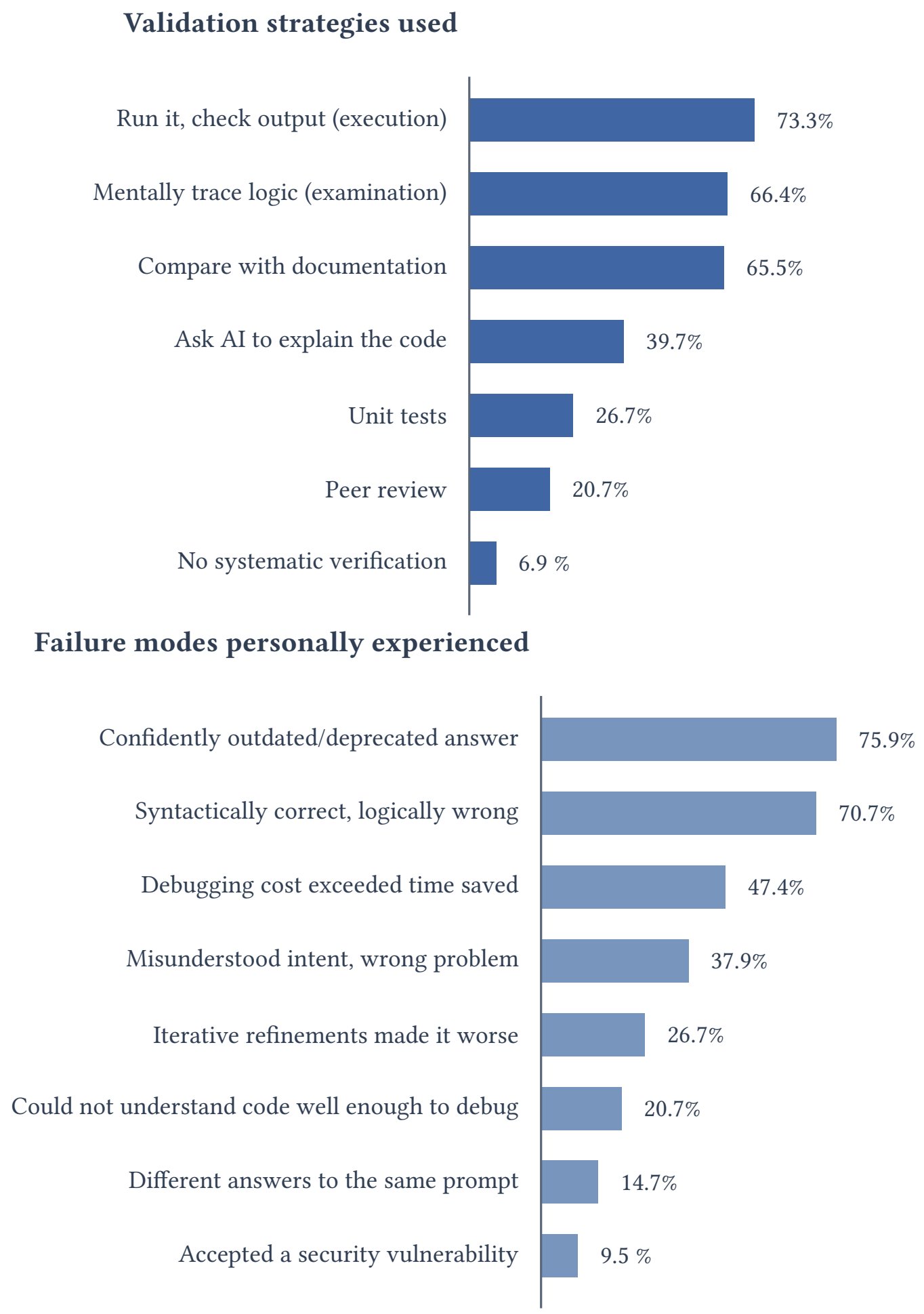

Only 25.0\% reported thoroughly reading every line of AI-generated code before running it. 44.8\% read but expect to miss subtle errors, and 30.2\% skim or run it immediately. Validation leaned heavily on execution. Running code and checking output (73.3\%) outranked mentally tracing logic (66.4\%), and only 26.7\% used unit tests and 20.7\% peer review (\autoref{fig:verification-failures}, left). Failure modes were nearly universal. 99.1\% had accepted AI code that later revealed a subtle bug at least once (78.5\% more than once or twice), 75.9\% had received confidently outdated answers, and 70.7\% had received syntactically correct but logically wrong code (\autoref{fig:verification-failures}, right).

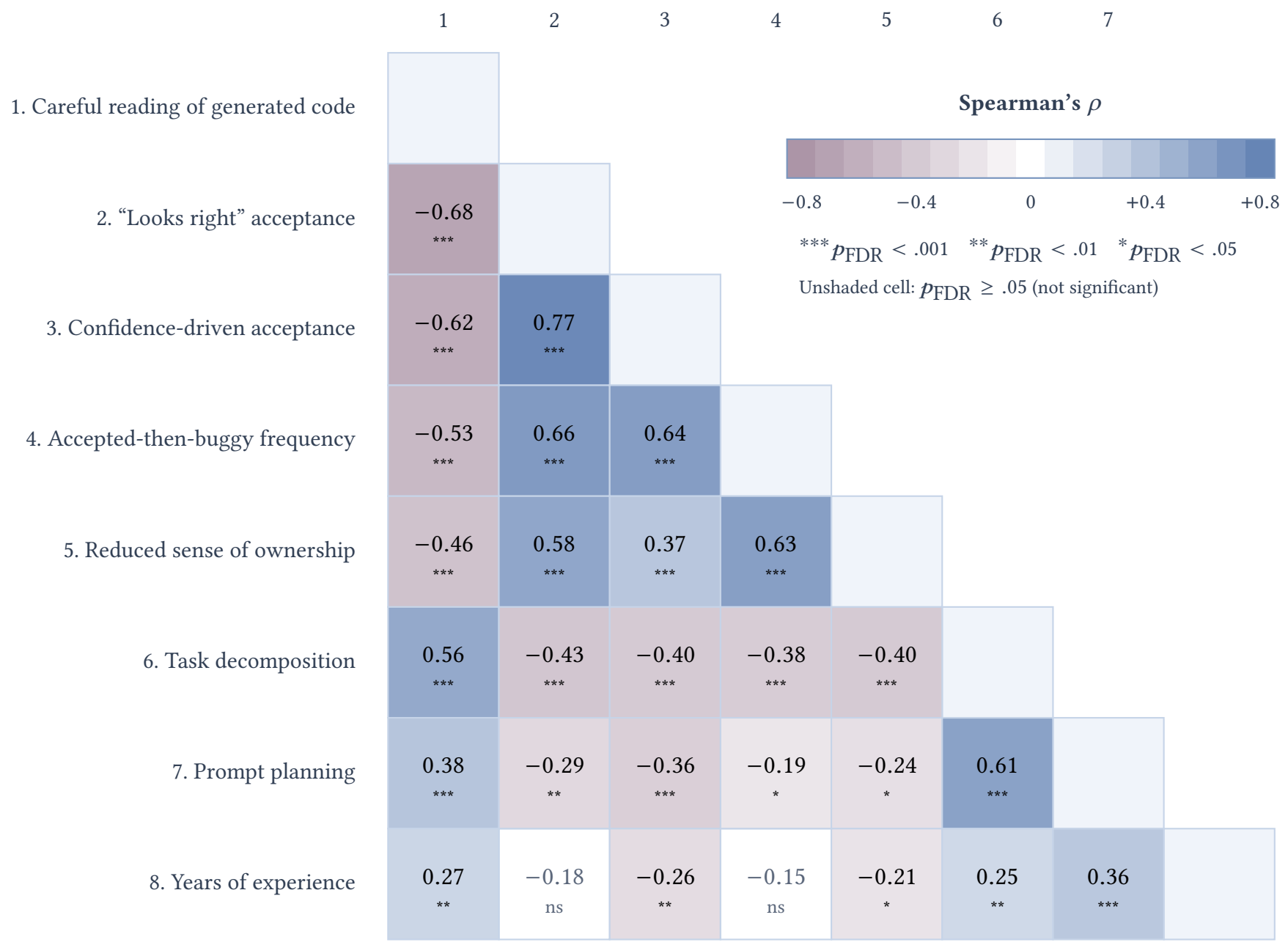
\begin{figure*}[t]
    \centering
    \begin{tikzpicture}[x=1.0cm, y=-1.0cm]
        \colorlet{poscorr}{paperBlue}
        \colorlet{negcorr}{paperNegative}

        \def\corrcell#1#2#3#4{%
            \pgfmathsetmacro{\pct}{abs(#3)*100}%
            \ifdim#3pt<0pt
                \fill[negcorr!\pct!white] (#1,#2) rectangle (#1+1,#2+1);
            \else
                \fill[poscorr!\pct!white] (#1,#2) rectangle (#1+1,#2+1);
            \fi
            \draw[paperRule, line width=0.4pt] (#1,#2) rectangle (#1+1,#2+1);
            \def\txtcol{black}
            \node[font=\scriptsize, text=\txtcol] at (#1+0.5,#2+0.43) {$#3$};
            \node[font=\tiny, text=\txtcol] at (#1+0.5,#2+0.72) {#4};
        }
        \def\nscell#1#2#3#4{%
            \draw[paperRule, line width=0.4pt] (#1,#2) rectangle (#1+1,#2+1);
            \node[font=\scriptsize, text=paperMuted] at (#1+0.5,#2+0.43) {$#3$};
            \node[font=\tiny, text=paperMuted] at (#1+0.5,#2+0.72) {#4};
        }

        \fill[paperWash] (0,0) rectangle (1,1);
        \draw[paperRule,line width=0.4pt] (0,0) rectangle (1,1);
        \fill[paperWash] (1,1) rectangle (2,2);
        \draw[paperRule,line width=0.4pt] (1,1) rectangle (2,2);
        \fill[paperWash] (2,2) rectangle (3,3);
        \draw[paperRule,line width=0.4pt] (2,2) rectangle (3,3);
        \fill[paperWash] (3,3) rectangle (4,4);
        \draw[paperRule,line width=0.4pt] (3,3) rectangle (4,4);
        \fill[paperWash] (4,4) rectangle (5,5);
        \draw[paperRule,line width=0.4pt] (4,4) rectangle (5,5);
        \fill[paperWash] (5,5) rectangle (6,6);
        \draw[paperRule,line width=0.4pt] (5,5) rectangle (6,6);
        \fill[paperWash] (6,6) rectangle (7,7);
        \draw[paperRule,line width=0.4pt] (6,6) rectangle (7,7);
        \fill[paperWash] (7,7) rectangle (8,8);
        \draw[paperRule,line width=0.4pt] (7,7) rectangle (8,8);

        \corrcell{0}{1}{-0.68}{***}
        \corrcell{0}{2}{-0.62}{***}
        \corrcell{1}{2}{0.77}{***}
        \corrcell{0}{3}{-0.53}{***}
        \corrcell{1}{3}{0.66}{***}
        \corrcell{2}{3}{0.64}{***}
        \corrcell{0}{4}{-0.46}{***}
        \corrcell{1}{4}{0.58}{***}
        \corrcell{2}{4}{0.37}{***}
        \corrcell{3}{4}{0.63}{***}
        \corrcell{0}{5}{0.56}{***}
        \corrcell{1}{5}{-0.43}{***}
        \corrcell{2}{5}{-0.40}{***}
        \corrcell{3}{5}{-0.38}{***}
        \corrcell{4}{5}{-0.40}{***}
        \corrcell{0}{6}{0.38}{***}
        \corrcell{1}{6}{-0.29}{**}
        \corrcell{2}{6}{-0.36}{***}
        \corrcell{3}{6}{-0.19}{*}
        \corrcell{4}{6}{-0.24}{*}
        \corrcell{5}{6}{0.61}{***}
        \corrcell{0}{7}{0.27}{**}
        \nscell{1}{7}{-0.18}{ns}
        \corrcell{2}{7}{-0.26}{**}
        \nscell{3}{7}{-0.15}{ns}
        \corrcell{4}{7}{-0.21}{*}
        \corrcell{5}{7}{0.25}{**}
        \corrcell{6}{7}{0.36}{***}

        \node[anchor=south, font=\scriptsize] at (0.5,-0.12) {1};
        \node[anchor=south, font=\scriptsize] at (1.5,-0.12) {2};
        \node[anchor=south, font=\scriptsize] at (2.5,-0.12) {3};
        \node[anchor=south, font=\scriptsize] at (3.5,-0.12) {4};
        \node[anchor=south, font=\scriptsize] at (4.5,-0.12) {5};
        \node[anchor=south, font=\scriptsize] at (5.5,-0.12) {6};
        \node[anchor=south, font=\scriptsize] at (6.5,-0.12) {7};

        \node[anchor=east, font=\scriptsize] at (-0.18,0.5) {1.~Careful reading of generated code};
        \node[anchor=east, font=\scriptsize] at (-0.18,1.5) {2.~``Looks right'' acceptance};
        \node[anchor=east, font=\scriptsize] at (-0.18,2.5) {3.~Confidence-driven acceptance};
        \node[anchor=east, font=\scriptsize] at (-0.18,3.5) {4.~Accepted-then-buggy frequency};
        \node[anchor=east, font=\scriptsize] at (-0.18,4.5) {5.~Reduced sense of ownership};
        \node[anchor=east, font=\scriptsize] at (-0.18,5.5) {6.~Task decomposition};
        \node[anchor=east, font=\scriptsize] at (-0.18,6.5) {7.~Prompt planning};
        \node[anchor=east, font=\scriptsize] at (-0.18,7.5) {8.~Years of experience};

        \node[anchor=south, font=\scriptsize\bfseries] at (5.80,0.68) {Spearman's $\rho$};
        \fill[negcorr!80!white] (3.6000,0.78) rectangle (3.8588,1.13);
        \fill[negcorr!70!white] (3.8588,0.78) rectangle (4.1176,1.13);
        \fill[negcorr!60!white] (4.1176,0.78) rectangle (4.3765,1.13);
        \fill[negcorr!50!white] (4.3765,0.78) rectangle (4.6353,1.13);
        \fill[negcorr!40!white] (4.6353,0.78) rectangle (4.8941,1.13);
        \fill[negcorr!30!white] (4.8941,0.78) rectangle (5.1529,1.13);
        \fill[negcorr!20!white] (5.1529,0.78) rectangle (5.4118,1.13);
        \fill[negcorr!10!white] (5.4118,0.78) rectangle (5.6706,1.13);
        \fill[poscorr!0!white] (5.6706,0.78) rectangle (5.9294,1.13);
        \fill[poscorr!10!white] (5.9294,0.78) rectangle (6.1882,1.13);
        \fill[poscorr!20!white] (6.1882,0.78) rectangle (6.4471,1.13);
        \fill[poscorr!30!white] (6.4471,0.78) rectangle (6.7059,1.13);
        \fill[poscorr!40!white] (6.7059,0.78) rectangle (6.9647,1.13);
        \fill[poscorr!50!white] (6.9647,0.78) rectangle (7.2235,1.13);
        \fill[poscorr!60!white] (7.2235,0.78) rectangle (7.4824,1.13);
        \fill[poscorr!70!white] (7.4824,0.78) rectangle (7.7412,1.13);
        \fill[poscorr!80!white] (7.7412,0.78) rectangle (8.0000,1.13);
        \draw[paperMid, line width=0.4pt] (3.60,0.78) rectangle (8.00,1.13);
        \node[anchor=north, font=\tiny] at (3.7294,1.17) {$-0.8$};
        \node[anchor=north, font=\tiny] at (4.7647,1.17) {$-0.4$};
        \node[anchor=north, font=\tiny] at (5.8000,1.17) {$0$};
        \node[anchor=north, font=\tiny] at (6.8353,1.17) {$+0.4$};
        \node[anchor=north, font=\tiny] at (7.8706,1.17) {$+0.8$};
        \node[anchor=west, font=\tiny] at (3.60,1.78) {$^{***}p_{\mathrm{FDR}}<.001$\quad $^{**}p_{\mathrm{FDR}}<.01$\quad $^{*}p_{\mathrm{FDR}}<.05$};
        \node[anchor=west, font=\tiny] at (3.60,2.14) {Unshaded cell: $p_{\mathrm{FDR}}\ge.05$ (not significant)};
    \end{tikzpicture}

    \caption{Spearman rank correlations among the eight verification, prompting, and experience items ($n=113$--$116$, pairwise deletion). Muted blue denotes a positive association and mauve a negative one; colour saturation is proportional to $|\rho|$, and cells that did not survive Benjamini--Hochberg correction across all 28 tests are left unshaded. Careful reading (item~1) is the most connected behaviour in the survey: it moves against every automation-bias and bad-outcome item and with prompting discipline and experience. Because these are same-respondent self-reports, the associations are exploratory rather than causal.}

    \Description{A lower-triangular correlation matrix of eight survey items, with the upper triangle left blank and used for the colour legend. Careful reading of generated code correlates negatively with looks-right acceptance at minus 0.68, with confidence-driven acceptance at minus 0.62, with accepted-then-buggy frequency at minus 0.53, and with reduced ownership at minus 0.46, and positively with task decomposition at 0.56, prompt planning at 0.38, and years of experience at 0.27. Looks-right acceptance and confidence-driven acceptance correlate at 0.77, and both correlate with accepted-then-buggy frequency at 0.66 and 0.64 respectively. Task decomposition and prompt planning correlate at 0.61. Years of experience shows the weakest associations overall; its correlations with looks-right acceptance at minus 0.18 and with accepted-then-buggy frequency at minus 0.15 are left unshaded because they are not significant after correction.}

    \label{fig:correlation-matrix}
\end{figure*}

How carefully a respondent read generated code was the single most connected behavior in the survey. Careful reading correlated negatively with accepting code because it ``looks right'' ($\rho=-.68$), with accepting confidently presented answers under less inspection ($\rho=-.62$), with accepted-then-buggy code frequency ($\rho=-.53$), and with reduced ownership feelings ($\rho=-.46$; all $p_{\text{FDR}}<.001$, $n=116$). ``Looks right'' acceptance and subtle-bug frequency rose together ($\rho=.66$, $p_{\text{FDR}}<.001$). Upstream prompting discipline showed the same direction at smaller magnitude. Task decomposition ($\rho=-.40$) and prompt planning ($\rho=-.36$) each correlated with less confidence-driven acceptance (both $p_{\text{FDR}}<.001$). \autoref{fig:correlation-matrix} lays out the full correlation structure across all eight items.

\subsubsection{Experience Shifts Verification Behavior}

\begin{figure}[t]
    \centering
    \begin{tikzpicture}[x=1.32cm, y=0.055cm]
        \colorlet{readbar}{paperBlue}

        \foreach \yv in {0,20,40,60} {
            \draw[paperRule, line width=0.4pt] (-0.15,\yv) -- (4.85,\yv);
            \node[anchor=east, font=\scriptsize, text=paperMuted] at (-0.2,\yv) {\yv\%};
        }

        \foreach \barlabel/\val/\n [count=\i from 0] in {
            {$<$1 yr}/10.0/10,
            {1--3 yrs}/12.9/31,
            {4--7 yrs}/9.7/31,
            {8--15 yrs}/50.0/26,
            {$>$15 yrs}/53.3/15
        } {
            \fill[readbar] (\i+0.08,0) rectangle (\i+0.62,\val);
            \node[anchor=south, font=\scriptsize] at (\i+0.35,\val) {\val\%};
            \node[anchor=north, font=\scriptsize, yshift=-2pt] at (\i+0.35,0) {\barlabel};
            \node[anchor=north, font=\tiny, text=paperMuted, yshift=-11pt] at (\i+0.35,0) {n=\n};
        }

        \draw[paperMuted,line width=0.6pt] (-0.15,0) -- (4.85,0);
    \end{tikzpicture}

    \caption{Share of respondents who report thoroughly reading every line of AI-generated code before running it, by years of programming experience (n=113; three respondents selecting ``Not applicable'' for experience are excluded). Thorough reading roughly quadruples past eight years of experience (Kruskal-Wallis $H=12.89$, $p=.012$).}

    \Description{A vertical bar chart with five experience bands. Thorough reading of AI-generated code is reported by 10 percent of developers with under one year of experience, 12.9 percent at one to three years, 9.7 percent at four to seven years, then jumps to 50 percent at eight to fifteen years and 53.3 percent above fifteen years.}

    \label{fig:reading-by-experience}
\end{figure}
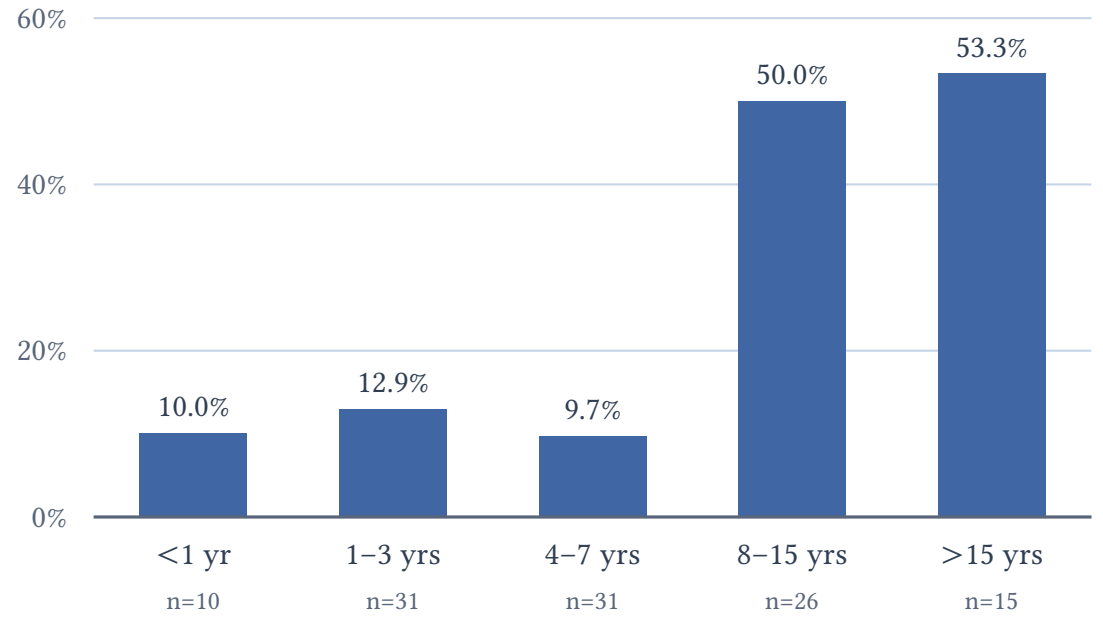

Years of experience correlated with more careful reading ($\rho=.27$, $p_{\text{FDR}}=.004$, $n=113$) and less confidence-driven acceptance ($\rho=-.26$, $p_{\text{FDR}}=.006$). The reading pattern was not a smooth gradient. The share reading every line thoroughly sat between 9.7\% and 12.9\% for every band up to seven years, then jumped to 50.0\% at 8--15 years and 53.3\% beyond (Kruskal-Wallis $H=12.89$, $p=.012$; \autoref{fig:reading-by-experience}). Respondents did not perceive AI as a junior-leveling tool. Asked who benefits most, 38.8\% named senior developers against 19.8\% naming juniors.

\subsubsection{Attitudes: Near-Universal Cognitive Debt, Strong Preference for Control}

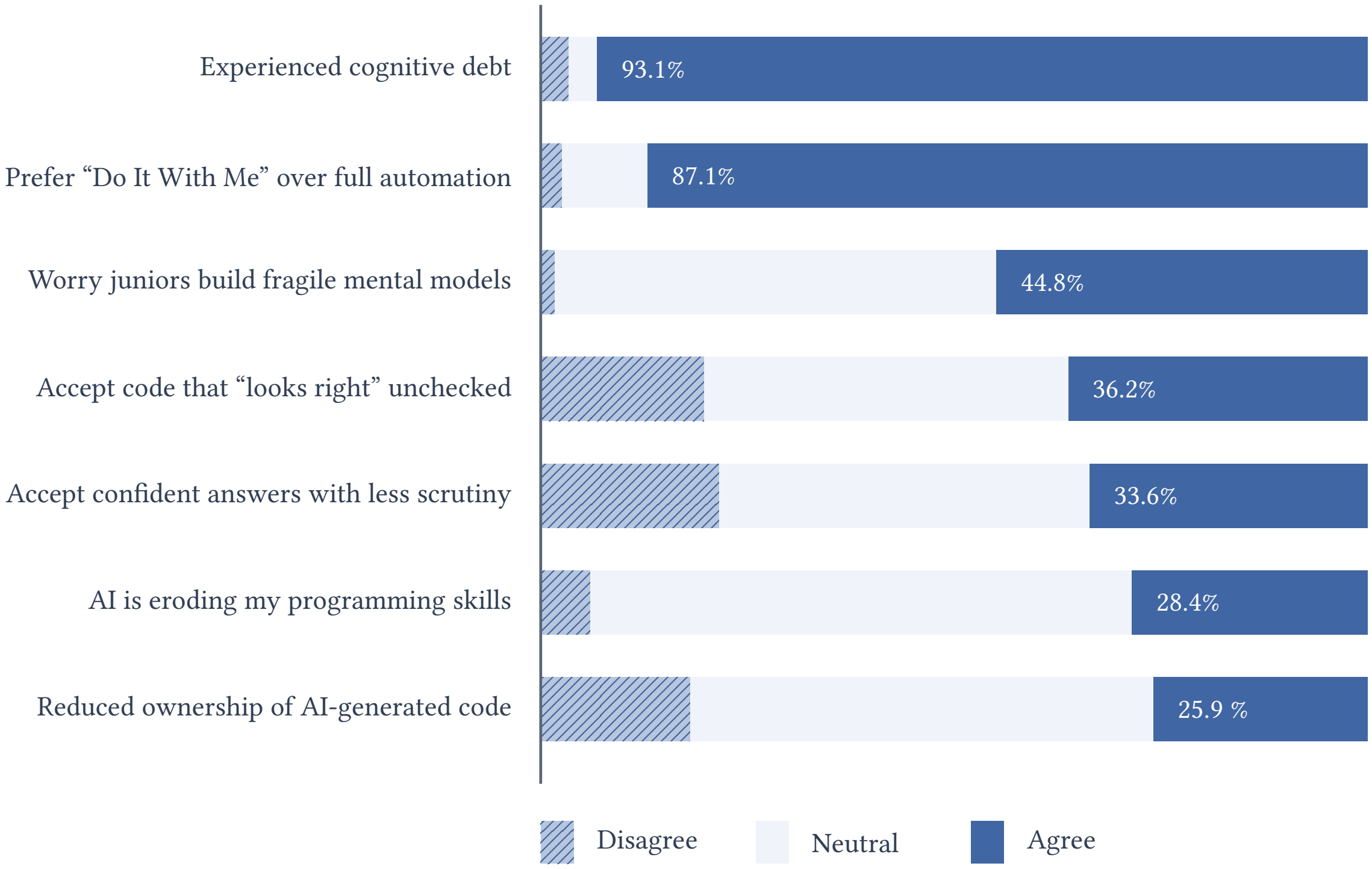
\begin{figure}[t]
    \centering
    \resizebox{\columnwidth}{!}{%
    \begin{tikzpicture}[x=0.062cm, y=-0.80cm]
        \colorlet{likdis}{paperLight}
        \colorlet{likneu}{paperWash}
        \colorlet{likagr}{paperBlue}

        \foreach \barlabel/\dis/\neu/\agr [count=\i from 0] in {
            {Experienced cognitive debt}/3.4/3.4/93.1,
            {Prefer ``Do It With Me'' over full automation}/2.6/10.3/87.1,
            {Worry juniors build fragile mental models}/1.7/53.4/44.8,
            {Accept code that ``looks right'' unchecked}/19.8/44.0/36.2,
            {Accept confident answers with less scrutiny}/21.6/44.8/33.6,
            {AI is eroding my programming skills}/6.0/65.5/28.4,
            {Reduced ownership of AI-generated code}/18.1/56.0/25.9
        } {
            \fill[likdis] (0,\i) rectangle (\dis,\i+0.6);
            \fill[pattern=north east lines, pattern color=paperBlue] (0,\i) rectangle (\dis,\i+0.6);
            \fill[likneu] (\dis,\i) rectangle ({\dis+\neu},\i+0.6);
            \fill[likagr] ({\dis+\neu},\i) rectangle (100,\i+0.6);
            \node[anchor=east, font=\scriptsize, xshift=-3pt] at (0,\i+0.3) {\barlabel};
            \node[anchor=west, font=\scriptsize, xshift=2pt, text=white] at ({\dis+\neu},\i+0.3) {\agr\%};
        }

        \draw[paperMuted,line width=0.6pt] (0,-0.3) -- (0,7.0);

        \fill[likdis] (0,7.35) rectangle (4,7.75);
        \fill[pattern=north east lines, pattern color=paperBlue] (0,7.35) rectangle (4,7.75);
        \node[anchor=west, font=\scriptsize] at (5,7.55) {Disagree};
        \fill[likneu] (26,7.35) rectangle (30,7.75);
        \node[anchor=west, font=\scriptsize] at (31,7.55) {Neutral};
        \fill[likagr] (52,7.35) rectangle (56,7.75);
        \node[anchor=west, font=\scriptsize] at (57,7.55) {Agree};
    \end{tikzpicture}%
    }

    \caption{Agreement with key single attitude items (N=116). Bars show the share of respondents disagreeing (strongly disagree + disagree), neutral, and agreeing (agree + strongly agree); the printed percentage is the agreeing share.}

    \Description{A stacked horizontal bar chart of seven Likert items. Agreement is near-universal for having experienced cognitive debt (93.1 percent) and preferring Do It With Me interaction (87.1 percent). 44.8 percent worry about juniors developing fragile mental models with almost no disagreement. Roughly a third agree they accept code that looks right without checking edge cases (36.2 percent) or accept confident answers with less scrutiny (33.6 percent). 28.4 percent agree AI is eroding their skills and 25.9 percent report reduced ownership of AI-generated code.}

    \label{fig:likert-attitudes}
\end{figure}

\autoref{fig:likert-attitudes} summarizes the key attitude items. Agreement that the respondent had personally experienced cognitive debt, meaning AI solved a problem they then did not understand, was near-universal at 93.1\%, and 87.1\% preferred ``Do It With Me'' over full automation. 44.8\% worried AI-heavy juniors will build fragile mental models (1.7\% disagreed). About a third endorsed the automation-bias behaviors above, and roughly a quarter reported reduced ownership (25.9\%) or felt their skills were eroding (28.4\%).

\subsubsection{Cost}

Of the 114 respondents for whom it applied, 64.9\% felt their AI spend was worth it. Exposure was often personal, with 20.7\% paying out of pocket, 19.8\% splitting costs with an employer, and only 34.5\% fully employer-covered. Personal spend concentrated at the low end (23.3\% at \$1--20, 15.5\% at \$21--50, 8.6\% above \$50), consistent with the cost sensitivity in \S\ref{sec:token-economy}.

\subsection{Qualitative Findings}
\label{sec:qual-findings}


The reflexive thematic analysis (\S\ref{sec:analysis}) organized 359 coded claims from the fourteen contextual inquiry sessions, their paired interviews, and 27 open-ended survey responses into seven themes and twenty-eight sub-themes. \autoref{tab:theme-participant-coverage} shows how unevenly they were distributed. The ownership probe (\S\ref{sec:ownership-comprehension}) surfaced in every session, while others appeared only where a participant's workflow made them relevant.





%
\begin{table*}[t]
\centering
\caption{Distribution of coded claims across contextual-inquiry participants (P1--P14), by theme and sub-theme. \CIRCLE\ = strong presence (5+ coded claims); \LEFTcircle\ = moderate presence (3--4); \Circle\ = light presence (1--2); blank = no coded claims from that participant in this sub-theme.}
\label{tab:theme-participant-coverage}
\tiny
\setlength{\tabcolsep}{3pt}
\begin{tabular}{@{}llcccccccccccccc@{}}
\toprule
Category & Sub-theme & \rotatebox{60}{P1} & \rotatebox{60}{P2} & \rotatebox{60}{P3} & \rotatebox{60}{P4} & \rotatebox{60}{P5} & \rotatebox{60}{P6} & \rotatebox{60}{P7} & \rotatebox{60}{P8} & \rotatebox{60}{P9} & \rotatebox{60}{P10} & \rotatebox{60}{P11} & \rotatebox{60}{P12} & \rotatebox{60}{P13} & \rotatebox{60}{P14} \\
\midrule
\multirow{4}{*}{\parbox{2.6cm}{Specification Labor}} & Plan as Artifact & \CIRCLE & \Circle & \Circle & \Circle & \LEFTcircle & \LEFTcircle & \LEFTcircle & \Circle & \LEFTcircle &  & \Circle & \CIRCLE & \LEFTcircle & \CIRCLE \\
 & Design Before Code &  &  &  &  &  & \CIRCLE &  &  &  &  &  &  &  & \CIRCLE \\
 & Requirement Translation & \Circle & \CIRCLE &  & \Circle &  & \LEFTcircle &  & \Circle & \Circle & \Circle & \Circle & \Circle & \LEFTcircle &  \\
 & The Planning Regret & \Circle &  &  & \Circle & \Circle & \Circle & \Circle &  & \Circle & \Circle & \Circle & \Circle &  & \Circle \\
\midrule
\multirow{4}{*}{\parbox{2.6cm}{Codified Convention}} & The Skills Layer & \Circle &  &  & \Circle & \Circle & \Circle & \Circle & \CIRCLE & \Circle & \LEFTcircle & \Circle & \Circle &  &  \\
 & Enforcing House Style &  &  & \LEFTcircle & \Circle &  &  &  &  &  & \Circle & \LEFTcircle & \LEFTcircle &  &  \\
 & Cheaper Substitutes for Skills &  &  &  &  &  &  &  & \Circle & \Circle &  &  &  &  &  \\
 & Where the Convention Layer Fails &  &  &  &  &  & \Circle &  & \Circle &  & \Circle &  &  &  &  \\
\midrule
\multirow{4}{*}{\parbox{2.6cm}{Token Economy}} & Sunk-Cost Acceptance & \Circle &  & \Circle &  & \Circle &  &  &  &  &  &  &  &  &  \\
 & Routing Work by Price &  & \LEFTcircle &  & \Circle & \Circle &  &  & \Circle &  &  &  &  &  &  \\
 & Doing It Yourself Is Cheaper &  &  & \Circle &  & \Circle & \Circle &  & \Circle & \Circle & \LEFTcircle & \Circle & \LEFTcircle & \LEFTcircle & \Circle \\
 & Felt Price Pressure &  & \Circle &  & \LEFTcircle &  &  & \Circle & \Circle &  & \Circle &  & \Circle & \Circle &  \\
\midrule
\multirow{4}{*}{\parbox{2.6cm}{Verification Burden}} & What Actually Gets Checked & \CIRCLE &  & \Circle & \Circle & \Circle & \Circle & \Circle & \Circle & \Circle & \LEFTcircle & \Circle & \Circle & \LEFTcircle & \LEFTcircle \\
 & QA as a Named Phase & \LEFTcircle &  &  & \Circle & \CIRCLE &  &  &  &  & \LEFTcircle &  & \Circle &  &  \\
 & When Verification Collapses &  & \CIRCLE &  & \Circle & \Circle & \Circle &  &  &  &  &  &  &  &  \\
 & Failures That Pass the Smoke Test &  &  & \Circle & \Circle & \Circle &  &  &  &  & \Circle &  &  &  & \Circle \\
\midrule
\multirow{4}{*}{\parbox{2.6cm}{What They Keep Doing Themselves}} & Setup and Dependencies & \CIRCLE &  & \LEFTcircle & \Circle &  &  &  &  & \LEFTcircle & \Circle & \Circle & \Circle & \Circle &  \\
 & Critical Features Kept Back &  & \Circle & \Circle & \Circle & \Circle &  & \Circle & \Circle & \Circle &  & \Circle & \Circle &  & \CIRCLE \\
 & Who Should Scaffold & \Circle &  & \Circle & \Circle &  &  & \LEFTcircle & \Circle &  & \Circle &  &  &  &  \\
 & Who Should Design &  &  &  &  & \Circle &  &  & \Circle & \Circle &  &  & \Circle &  & \Circle \\
\midrule
\multirow{4}{*}{\parbox{2.6cm}{Ownership \& Comprehension}} & The Ownership Probe &  & \Circle & \Circle & \Circle & \Circle & \Circle &  & \Circle & \Circle & \Circle & \Circle & \Circle & \Circle & \Circle \\
 & Routes to Ownership & \Circle &  & \Circle & \Circle &  &  & \Circle & \Circle & \Circle & \Circle & \LEFTcircle & \Circle &  & \Circle \\
 & Comprehension Loss & \LEFTcircle & \Circle &  & \Circle & \Circle & \Circle & \Circle &  &  &  &  &  & \Circle &  \\
 & Building the Mental Map on Purpose & \Circle &  &  &  &  &  &  &  &  &  & \LEFTcircle & \Circle &  &  \\
\midrule
\multirow{4}{*}{\parbox{2.6cm}{AI Skill as Its Own Competence}} & Experience vs AI fluency &  &  &  &  &  &  & \Circle & \LEFTcircle &  &  &  &  & \Circle &  \\
 & Fundamentals as Precondition & \Circle & \Circle & \LEFTcircle &  &  & \Circle & \Circle & \Circle &  & \Circle &  &  &  &  \\
 & Learning With AI &  & \LEFTcircle &  &  &  & \Circle &  &  & \Circle &  & \Circle &  &  &  \\
 & Prompting as a Hireable Skill &  &  &  &  &  &  &  & \Circle &  &  &  &  &  &  \\
\bottomrule
\end{tabular}
\end{table*}

\subsubsection{Specification Labor}
\label{sec:specification-labor}

The plan routinely stopped being a chat message and became a file in the repository. Participants used PLAN.md, requirements.md, or an implementation\_plan.md that the agent produced, and the participant moved into the project folder. P10 described a full pipeline built on it.

\begin{quote}
``If it's a very large project, we have a PLAN.md where all the steps are written. Each feature goes under a new branch, then code generation, then unit testing, then a PR. All these steps are usually done by Claude. After my manual QA it gets merged into the main codebase.'' --- P10
\end{quote}

Several participants cut the plan into phases the agent had to stop and wait for approval on, and P1 said he opens with a planning prompt ``all the time.'' Frontend work got parallel treatment. Participants imported DESIGN.md files from Stitch, Figma, or getdesign.md, or sketched the interface in Excalidraw first, because ``if they are not created beforehand, AI gives bad output'' (P5). Asked what they would do differently with more time, the subset who answered named more planning, better prompts, and more QA. None mentioned writing more code themselves.

\subsubsection{Codified Convention}
\label{sec:codified-convention}

Several participants described maintaining a shared skill installed on every team machine so the agent writes in-house style. P8 narrated the transition.

\begin{quote}
``The issue we faced back then was that every team member had their own .md files, and we had to add project-specific files to keep everything on the same page. Then we learned that we can organize these .md files and create skills ourselves. So we shifted to skills and actively use them in daily work.'' --- P8
\end{quote}

P4 framed the need from a team without it. ``There is a thing called SKILLs in AI development [...] common SKILL files for instruction are much needed there.'' Some skills were narrower, bound to a single job rather than a whole project, such as P8's \texttt{submit-the-pr} skill or the UI/UX skill P10 called from skill.sh mid-session. Not everyone reached for one. Some enforced house style by hand, writing one file and telling the agent to copy its pattern (P3, P11, P12). P8's team defaulted to templates instead, reasoning that ``a template doesn't require any AI power, so no token use [...] skills are nothing but instructions written down, and when calling them AI loads and thinks every time, so there is a cost associated with them.'' Off-the-shelf skills pulled from public sources fared worst. SP16 tried skills.sh and got ``wrong folder layout, unfamiliar component style, and inconsistent naming.'' The same respondent's custom skills, which encoded their own folder structure and review standards, reportedly produced a tenfold jump in a colleague's velocity. We treat that single secondhand figure as illustrative rather than measured.

\subsubsection{Token Economy}
\label{sec:token-economy}

Cost operated as a mechanism inside the loop, not as background context. Participants accepted doubted output because discarding it would waste tokens already spent.

\begin{quote}
``I'm accepting because this code generation requires token cost. If I abandon this output, all the tokens will be wasted. I'll fix the code later if there is any bug.'' --- P1
\end{quote}

The same logic governed which model got used for which job. P8 routed cheap models to routine generation and reserved expensive ones for verification.

\begin{quote}
``We should use a powerful model for a powerful task. In our team, many people use very powerful models for very basic tasks. But powerful models have a high token price, so there is waste. In our company, most of the tasks can be done with Sonnet. If we need thorough verification of something, we go with Opus.'' --- P8
\end{quote}

Doing a fix by hand was consistently framed as arithmetic rather than distrust. ``In the time I spend giving it instructions, I can complete the fix myself. Why waste tokens for that small issue?'' (P12). P7 named cost as his single greatest difficulty.

\subsubsection{Verification Burden}
\label{sec:verification-burden}
The second-largest theme describes where the time AI saves actually goes: into checking and into deciding what to check. Some participants read every generated file before running anything (P4, P5, P14). Others verified almost entirely through the rendered interface and left the source unread (P6, P13). P13 illustrated a riskier pattern. He stopped the agent mid-run when its reasoning went wrong, but could not locate a small bug in the finished code afterward. Several promoted verification into a named phase, splitting sessions into planning, implementation, and QA (P5) or running generated code through unit tests and a pull request before manual QA (P10). Verification also collapsed under specific conditions.

\begin{quote}
``The model explained its approach so confidently, with a tidy summary of edge cases it claimed to handle, that I merged the change after a skim [...] The confident tone did the reviewing for me, and that's exactly the trap.'' --- SP20
\end{quote}

The characteristic failure was output that looked right, compiled, and passed a quick demo, then broke somewhere the demo did not reach. Examples included contract drift between a Go backend and its TypeScript frontend (P14), a fix that silently disabled input validation (SP26), and quietly out-of-date packages (P1). None announced itself with a crash.

\subsubsection{What They Keep Doing Themselves}
\label{sec:retained-half}

No two participants drew the delegation boundary in the same place. The most common retained task was project setup, done by hand out of distrust in package choices.

\begin{quote}
``Claude sometimes installs a version that doesn't even exist. Even if that exists, sometimes that is backdated, which often breaks the code later. That's why I usually initialize the project manually and install the up-to-date packages.'' --- P1
\end{quote}

Authentication, data models, and schema design were named repeatedly as work too expensive to get wrong. P9 said, ``I go solo when doing critical tasks like creating models or auth features.'' P14 kept an entire layer, writing the Go backend by hand while letting AI shape the frontend, and called the split ``the ideal developer sweet spot right now.'' Scaffolding produced the sharpest contradiction. P7 named it the task AI suits best, then spent longer reorganizing AI-generated folders than those who had scaffolded by hand (P3, P12). Frontend design split similarly, weak without a guideline (P8 called AI ``dumb'' in the frontend) and strong when working from an imported one (P14).

\subsubsection{Ownership and Comprehension}
\label{sec:ownership-comprehension}

A structured probe asked participants to explain, fix, or change a part of the application they had just built. Response speed tracked what the participant had written or deliberately read, not job title or experience. Participants who had written the critical code went straight to the right file (P3, P11, P12). Others could not locate it at all (P2, P6, P13). A second route ran through convention rather than authorship. P8 and P10 had barely typed any of their session's code, but a global team skill had forced the agent into structures they already knew, so both moved as fast as participants who had written everything by hand. P11 reached the same place a third way, reading every generated function name on purpose. Where none of these routes were present, comprehension thinned quickly.

\begin{quote}
``In my first month I pushed code that the AI basically wrote for me, and in code review I couldn't answer why the code was structured that way. It worked, but I didn't understand it, and I had to go back and relearn the thing I had supposedly built.'' --- SP19
\end{quote}

\subsubsection{AI Skill as Its Own Competence}
\label{sec:ai-skill-competence}

Fluency with AI-assisted workflows behaved as a skill in its own right, separate from general engineering experience. P7, fourteen years into his career, described his own AI literacy candidly. ``I can feel I don't have much AI literacy.'' SP18, with over twenty years of experience, called an early agent session ``humbling.'' P8 described the gap running the other way. ``Seniors are less likely to adopt AI, and juniors or freshers adopt AI faster [...] Juniors don't understand the edge cases, which is where AI is very bad [...]'' Across both directions, the most repeated answer to what separates a good AI-assisted developer from a poor one was understanding of fundamentals.

\begin{quote}
``If you don't understand what AI is generating, you can't integrate it into your system [...] AI-assisted development is a paradigm shift, but fundamentally development remains the same.'' --- P3
\end{quote}

At the far end of this theme, P8 described prompt quality becoming a formal hiring criterion. ``[W]e now hire people by examining their prompt quality and prompt optimization technique.''

\subsection{Triangulating the Quantitative and Qualitative Themes}
\label{sec:findings-triangulation}

Following mixed-methods integration practice \cite{Fetters2013Integration, OCathain2010Three}, we join the two strands theme by theme in a joint display (\autoref{tab:joint-display}) rather than gesturing at overlap afterward. For each of the seven qualitative themes, we asked whether an independently measured quantitative signal exists, and if so, whether it agrees, adds something the qualitative record could not see, or pulls in a different direction. We treat all four answers as findings, namely agreement, complementary expansion, tension, and outright silence. A theme the instrument could not see is evidence about its scope, not a hole in the analysis. 

\definecolor{patternconverge}{HTML}{2E7D32}
\definecolor{patterncomplement}{HTML}{3D7EA6}
\definecolor{patterntension}{HTML}{C97B1E}
\definecolor{patternsilent}{HTML}{6B6B6B}
\newcolumntype{L}[1]{>{\raggedright\arraybackslash}p{#1}}
\newcolumntype{C}[1]{>{\centering\arraybackslash}p{#1}}

\begin{table*}[t]
\centering
\footnotesize
\setlength{\tabcolsep}{4pt}
\renewcommand{\arraystretch}{1.22}
\caption{Joint display triangulating the seven qualitative themes against the quantitative strand, following mixed-methods integration practice \cite{Fetters2013Integration, OCathain2010Three}. \textcolor{patternconverge}{\textbf{Convergent}}: an independently measured quantitative signal agrees with the qualitative pattern. \textcolor{patterncomplement}{\textbf{Complementary}}: the quantitative item establishes prevalence while the qualitative record supplies a mechanism the item cannot measure. \textcolor{patterntension}{\textbf{Tension}}: an aggregate quantitative trend coexists with qualitative counter-cases at the individual level. \textcolor{patternsilent}{\textbf{Silent}}: no survey item corresponds to the theme; we report this as a finding about the instrument's scope, not a gap in the analysis.}
\label{tab:joint-display}
\begin{tabular}{@{}L{1.8cm}L{3.25cm}L{3.45cm}C{2.25cm}L{2.85cm}@{}}
\toprule
\textbf{Theme} & \textbf{Qualitative Evidence} & \textbf{Quantitative Signal} & \textbf{Pattern} & \textbf{What It Means} \\
\midrule

Specification Labor &
Plans leave the chat and become a repository file (PLAN.md, requirements.md); design files imported before frontend code (\S\ref{sec:specification-labor}) &
60.3\% usually/almost always plan a prompt; 82.8\% often decompose tasks; both correlate with less confidence-driven acceptance ($\rho=-.36$, $-.40$) &
\textcolor{patternconverge}{\textbf{Convergent}} &
Discipline that is protective at sample scale takes a concrete, observable form on the ground: a file, not a mood. \\
\addlinespace

Codified Convention &
Teams install a shared skill or house-style file so agent output stays uniform; some substitute templates instead (\S\ref{sec:codified-convention}) &
No survey item; nearest proxy (a desired-improvement checkbox for project-context awareness) measures a wish, not current practice &
\textcolor{patternsilent}{\textbf{Silent}} &
The instrument could not see a practice that only becomes visible by watching people work, not by asking them. \\
\addlinespace

Token Economy &
Cost drives sunk-cost acceptance, model routing by task, and the threshold for a manual fix (\S\ref{sec:token-economy}) &
64.9\% (of 114) felt spend worth it; 20.7\% pay personally, 19.8\% split with an employer; personal spend concentrated at \$1--20 (23.3\%) &
\textcolor{patterncomplement}{\textbf{Complementary}} &
Quant maps how widely the cost burden is shared; only the sessions show the accept-anyway logic it produces. \\
\addlinespace

Verification Burden &
Second-largest theme in the corpus; named QA phases, a confidence trap (SP20), failures that pass a working demo (\S\ref{sec:verification-burden}) &
Careful reading is the most connected item in the survey (\autoref{fig:correlation-matrix}); 99.1\% accepted buggy output at least once &
\textcolor{patternconverge}{\textbf{Convergent}} &
The survey's strongest statistical signal and the corpus's second-largest theme are the same finding, seen twice. \\
\addlinespace

What They Keep Doing Themselves &
Every participant drew a different delegation boundary; setup, dependencies, auth, and schema design were kept back most often (\S\ref{sec:retained-half}) &
Q7 ranks nine generic delegated tasks; none names setup, dependencies, auth, or schema, so retention has no matching item &
\textcolor{patternsilent}{\textbf{Silent}} &
An instrument built to measure what gets delegated most cannot, by construction, measure what gets refused. \\
\addlinespace

Ownership \& Comprehension &
Comprehension tracked authorship or deliberate reading; a shared convention substituted for both for P8 and P10 (\S\ref{sec:ownership-comprehension}) &
Reduced-ownership item (25.9\% agree) correlates with less careful reading ($\rho=-.46$, $p_{\text{FDR}}<.001$) &
\textcolor{patterncomplement}{\textbf{Complementary}} &
Quant confirms the one route to ownership a single item can ask about; the sessions find a second. \\
\addlinespace

AI Skill as Its Own Competence &
P7 (14 years) called his own AI literacy thin; SP18 (20+ years) called an early agent session humbling (\S\ref{sec:ai-skill-competence}) &
Experience correlates modestly with careful reading ($\rho=.27$); thorough reading jumps sharply at 8+ years, not smoothly (\autoref{fig:reading-by-experience}) &
\textcolor{patterntension}{\textbf{Tension}} &
An aggregate trend and its individual exceptions are both true at once; that gap is the competence paradox (\S\ref{sec:competence-paradox}). \\

\bottomrule
\end{tabular}
\end{table*}

\subsubsection{Convergent: Where the Strands Agree}


Two themes are convergent in the strict sense, meaning an independently measured quantitative pattern lines up with what the sessions showed directly. Specification labor is the first. The survey's planning and decomposition items (60.3\% usually or almost always plan a prompt, 82.8\% often decompose tasks) both correlate with less confidence-driven acceptance (\S\ref{sec:quant-findings}), and the sessions show what that discipline looks like. A plan leaves the chat window and becomes PLAN.md, a phase gate, or a design file imported before code (\S\ref{sec:specification-labor}). Verification burden is the second and strongest. Careful reading is the most connected variable in the correlation structure (\autoref{fig:correlation-matrix}). The survey's sharpest pattern is that ``looks right'' acceptance and later-discovered bugs rise together while careful reading suppresses both. That is precisely the failure the sessions caught live, in SP20's confidence trap and in P13 stopping a run mid-reasoning and then being unable to locate the bug that remained (\S\ref{sec:verification-burden}). Between them, these two themes carry 159 of the corpus's 359 coded claims and touch all fourteen contextual-inquiry participants. The strongest qualitative material is also the best corroborated quantitatively.

\subsubsection{Complementary: Where the Strands Answer Different Questions}

Two more themes are corroborated, but the quantitative item establishes prevalence while the qualitative record supplies a mechanism, a pattern mixed-methods researchers distinguish from strict convergence because neither strand alone would produce the finding \cite{Fetters2013Integration}. Token economy is descriptively well supported. 64.9\% of the 114 respondents for whom it applied felt their AI spend was worth it, and personal exposure was common (20.7\% paid alone, 19.8\% split with an employer). What the survey cannot show is why a participant would accept output they doubted. That comes only from P1's sunk-cost arithmetic and P8's routing of cheap models to routine work and expensive ones to verification (\S\ref{sec:token-economy}). Ownership and comprehension work the same way from the other direction. The reduced-ownership item correlates with less careful reading exactly as the sessions would predict ($\rho=-.46$), but cannot distinguish authorship from the second route the sessions revealed: comprehension carried by shared convention rather than by typing the code, which let P8 and P10 move as fast as participants who had written everything themselves (\S\ref{sec:ownership-comprehension}). In both cases, the quantitative strand confirms the part of the mechanism it was built to detect and is silent on the rest, a more productive silence than the one below.

\subsubsection{Tension: Where Aggregate and Individual Disagree}

AI skill as its own competence produces the study's clearest tension, not between the strands' conclusions but between their grain. At the sample level, experience is doing real work. Thorough reading holds flat across the first seven years before jumping sharply at eight, and years of experience correlates modestly with more careful reading and less confidence-driven acceptance (\autoref{fig:reading-by-experience}; \S\ref{sec:quant-findings}). Read at the individual level, the same theme contains its own counter-cases. P7, fourteen years into his career, described his own AI literacy as thin, and SP18, with over twenty years of experience, called an early agent session ``humbling'' (\S\ref{sec:ai-skill-competence}). Both readings are correct at once. A trend that holds in aggregate does not guarantee that any particular senior developer has the tacit skill it describes, which is the paradox this paper is named for (\S\ref{sec:competence-paradox}).

\subsubsection{Silent: Where the Instrument Had Nothing to Say}

Two themes have no corresponding survey item, which we report as a finding about the instrument's reach. Codified convention, the shared skill or house-style file a team installs on every machine, was never asked about directly. The nearest item, a desired-improvement checkbox for ``better awareness of project context,'' measures a wish, not a practice. Silence in the survey did not make the theme thin. It carries 46 coded claims from eleven of fourteen participants and six survey free-text responses, and holds its own internal tension independent of any quantitative check. P12 and P6 experienced the instruction layer as unpaid overhead, and P8's team chose templates over skills for exactly that reason. Against them sits SP16's secondhand account of a tenfold velocity gain from a skill that encoded a team's actual conventions rather than a generic one (\S\ref{sec:codified-convention}). What they keep doing themselves is silent for a more structural reason. Q7 asked which of nine generic tasks respondents delegate most, rather than what they withhold, and none of the nine corresponds to the work the sessions show being kept back, namely environment setup, dependency choices, authentication, and schema design (\S\ref{sec:retained-half}). An instrument built around delegation frequency cannot see a boundary defined by refusal. Neither theme is weaker for lacking quantitative corroboration. Both report on territory the survey was never built to cover, and fourteen independent sessions converging there, particularly on the persistent-instructions need in \S\ref{sec:codified-convention}, is itself evidence that the gap is real rather than a design oversight.

\section{Discussion} \label{sec:discussion}

This section reads the two strands of \S\ref{sec:findings} against each other, using the joint display in \S\ref{sec:findings-triangulation} (\autoref{tab:joint-display}). Where the strands agree, we treat the finding as settled within the limits of the sample. Where one is silent, we treat that silence as a boundary of the instrument rather than of the phenomenon.

\subsection{The Competence Paradox of Vibe Coding}
\label{sec:competence-paradox}

Vibe coding is typically pitched as a leveling technology. Natural language stands in for syntax, and the distance between idea and prototype collapses for novice and expert alike \cite{Li2026VibeCoding, Karpathy2025VibeCoding}. Our data complicate this. Self-reports linked lower automation-bias endorsement to careful reading, prompt planning, and task decomposition (\S\ref{sec:quant-findings}). These are exactly the disciplines the sessions show taking real, unpaid effort to build, such as a PLAN.md maintained across features (\S\ref{sec:specification-labor}), a shared skill installed on every machine (\S\ref{sec:codified-convention}), and a named QA phase before merge (\S\ref{sec:verification-burden}). Nobody arrives at these habits passively, and they do not track seniority simply. Thorough reading stays flat across the first seven years of experience before jumping sharply (\autoref{fig:reading-by-experience}). The qualitative record runs both directions. A developer fourteen years into his career called his own AI literacy thin, and another with over twenty years of experience called an early agent session ``humbling'' (\S\ref{sec:ai-skill-competence}). Both point to a distinct, largely implicit competency that some senior engineers have not yet acquired and some juniors already have.

Two threads in the literature anticipated this without quite stating it. The first is delegation calibrated by task familiarity rather than tenure \cite{Feng2026AgencyAI}. The second is natural-language specification treated as a skill to be taught, not assumed \cite{Zamfirescu2023WhyJohnny, Subramonyam2025Prototyping}. What our findings add is a mechanism. The tool meant to flatten the expertise curve instead rewards a second, mostly tacit expertise in operating it. Persistent instruction files are the artifact meant to transfer that expertise across a team, and by the only large-scale study available, they are badly authored more often than not \cite{Hong2026SkillSmells}. The paradox is not that vibe coding fails novices outright. It is that the version that actually works looks like ordinary engineering discipline running through a conversational interface, and it is currently available mostly to people who already had that discipline.

Our population adds a specific inflection worth naming. Interaction with these tools happens almost entirely in English, which is not the first language of any of our participants. Work on language-model performance across users finds that non-native English speakers receive lower-quality responses even when both write in English \cite{Reusens2024NativeBias}, and prompting research already treats articulation quality as a major driver of output quality \cite{Wang2024UML, Pickering2025HowHumans}. We did not measure English proficiency directly, so we cannot separate its contribution from the under-specification the literature already documents. The mechanism is plausible enough to belong on the research agenda.

\subsection{Verification Displaces Generation, It Does Not Replace It}
\label{sec:verification-displaces}

Verification is the second-largest theme in our qualitative corpus (\S\ref{sec:verification-burden}), and the survey's strongest associations sit exactly here. Self-reported reading care correlated with every bad outcome we measured. At the same time, the upstream disciplines of prompt planning and task decomposition showed the same direction at smaller magnitude (\S\ref{sec:quant-findings}). These are same-respondent self-report items, so part of the correlation could reflect general conscientiousness rather than reading care itself. We read them as consistent with the qualitative mechanism, not proof of it. With that caution in mind, reading is the verification act itself, and the upstream disciplines make it tractable once the code arrives. This lines up with the distinction between execution-based and examination-based validation \cite{Ferdowsi2024Validating}. The failures in our data that survived a working demo were out-of-date packages, contract drift between a backend and its frontend, and a fix that silently disabled input validation. These were exactly the ones execution alone could not surface, and they were caught only by participants who had budgeted time to read the code. SP20's account of being talked out of a careful review by a confidently worded summary is the clearest instance of the automation bias the survey measures \cite{Goddard2012AutomationBias, Li2025From}. It happened to an experienced respondent, which argues against treating automation bias as a novice-only failure.

The most common item on the kept-manual side of the delegation boundary was project initialization and dependency management (\S\ref{sec:retained-half}). This reads as a rational response to the same asymmetry rather than general distrust. A wrong package choice is expensive to detect after the fact and cheap to prevent by doing it manually. Participants were pricing in where a missed check costs most, not withholding trust across the board. Tooling that surfaces this asymmetry explicitly, flagging changes most likely to fail silently rather than treating all output as equally reviewable, would extend what our participants were already doing by hand.

\subsection{Cost as a First-Class Design Constraint}
\label{sec:cost-constraint}

Token cost shaped which output got accepted, which model got assigned to which job, and when a participant abandoned the agent for a fix by hand (\S\ref{sec:token-economy}). P1 accepted doubted output because the tokens were already spent, and P8 routed routine generation to cheaper models while reserving expensive ones for verification. In both cases, cost operates as a variable inside the loop, not as a constraint external to it. This sits next to a growing systems literature on reducing agent-trajectory cost \cite{Xiao2026TrajectoryReduction, Zhang2022DietCode} and vision-level calls for AI collaborators that manage cost on a developer's behalf \cite{Hassan2026SE3}. Nothing in the tools our participants used exposed cost as a comparative, in-context signal. They estimated it from provider dashboards and billing cycles, as in SP27's end-of-cycle downgrade to a cheaper model.

The gap matters because cost pressure interacts directly with the verification problem above. Accepting code because it has already been paid for, rather than because it is trusted, is a distinct failure mode from the confidence-driven automation bias the trust literature usually describes \cite{Li2025From}. A developer can doubt the code and accept it anyway, for reasons that have nothing to do with trust.

\subsection{Persistent Context Artifacts: A Fix Whose Own Quality Is Unverified}
\label{sec:context-artifacts-gap}

Participants converged, independently and without prompting from our interview guide, on the same response to the fragility of one-off prompting. They moved project convention out of the conversation and into a file the agent loads automatically (\S\ref{sec:codified-convention}). No survey item corroborates this mechanism. It belongs to the silent cell of \autoref{tab:joint-display}, and rests on the qualitative record alone, on fourteen sessions independently arriving at the same fix for an instrument that never asked about it. Where this worked, it did double duty. It stabilized output style, and for P8 and P10 it substituted for authorship-based comprehension (\S\ref{sec:ownership-comprehension}), because a team skill had shaped the agent's output into a form they already recognized. This sits awkwardly next to a cognitive debt literature that treats comprehension as producible only by direct authorship or deliberate reading \cite{Kosmyna2025CognitiveDebt, Ali2025EvenThough}. Convention, on our evidence, is a third route to the same place.

The same theme contains its own counter-evidence. SP16's off-the-shelf skill produced worse output than no skill at all, while a custom skill encoding that team's actual conventions reportedly produced a tenfold velocity gain for a colleague. That is a single secondhand account, suggestive rather than measured. The direction, at least, tracks authoring quality rather than a skill's mere presence. This is the gap the one systematic study of the artifact has since documented at scale. Over 99\% of 238 real SKILL.md files contained at least one authoring smell, and smells introduced early rarely got corrected \cite{Hong2026SkillSmells}. Our participants were discovering by trial and error a quality problem the literature has only begun to name. P8's team chose GitHub templates over skills because ``a template doesn't require any AI power, so no token use.'' That reads less as rejection of the idea than as a rational response to an artifact whose quality nobody could yet guarantee.

Read together, these four threads describe a single condition rather than four independent findings. AI-assisted development works when someone rebuilds, largely by hand, the specification, verification, cost-accounting, and shared convention that a conversational interface does not supply on its own. It works badly when no one does. Nothing in our data suggests this scaffolding is optional, or specific to our population. It looks like the ordinary discipline of software engineering, reappearing wherever the interface has stripped away its usual visible cues. Our data show that only some developers reliably reconstruct this discipline on their own. The question is therefore not whether they can be trusted to do it. It is how tooling can make disciplined use the default rather than the exception that a minority happen to rebuild. \S\ref{sec:design-implications-recommendations} takes that up directly.

\section{Design Implications / Recommendations} \label{sec:design-implications-recommendations}
We turned \S\ref{sec:discussion} into a design brief of five implications, each anchored to a specific finding and stated as something a tool or organization could do differently. We take the third furthest, treating instruction files as software artifacts. \S\ref{sec:better-skills} describes Better Skills, a working design probe built on it.

\subsection{Make Verification Legible, Not Just Possible}
\label{sec:impl-verification}

Verification was the second-largest theme in our qualitative corpus and the survey's strongest set of associations (\S\ref{sec:verification-displaces}), but the tools our participants used treated every change as equally reviewable. Our data say otherwise. The failures that mattered were out-of-date packages, contract drift between services, and a validation check silently disabled. Each passed a working demo and was caught only by participants who had budgeted time to read rather than run the code. So, flag dependency changes, cross-service contracts, and security-relevant edits as requiring examination rather than execution-based trust, since these were exactly the categories where our participants' checks failed silently. Surface a change's failure-visibility class next to the diff itself, not in a separate lint report, so the decision to skim or read closely is informed rather than habitual.

\subsection{Expose Cost as a Comparative, In-Context Signal}
\label{sec:impl-cost}

Token cost shaped which output got accepted, which model ran which job, and when a participant abandoned the agent for a fix by hand (\S\ref{sec:cost-constraint}), yet nothing in the tools themselves exposed cost as a signal. Participants estimated it from provider dashboards and billing cycles, disconnected from the moment of acceptance. Thus, show token cost next to the model's output, not only in a billing dashboard, so decisions like P8's model routing or P12's do-it-by-hand arithmetic can draw on real numbers rather than felt estimates. Distinguish ``already paid for'' from ``verified'' in the interface, since our data show these two get conflated into a single acceptance decision.

\subsection{Treat Instruction Files as Software Artifacts, Not Text}
\label{sec:impl-instruction-files}

Persistent skill and rules files are becoming the default way of encoding team convention (\S\ref{sec:context-artifacts-gap}). If so, they need the quality infrastructure code already has, and the authors with the judgment to write them well. Neither exists reliably today. The one large-scale study of the artifact found over 99\% of real-world SKILL.md files carried at least one authoring smell \cite{Hong2026SkillSmells}, and our own participants improvised these files without any shared standard for what a good one looks like. Lint skill and rules files against known authoring smells \cite{Hong2026SkillSmells}, version them, and route them through the same review process as the code they govern. Recognize skill authoring as senior engineering work rather than a clerical task. The judgment a good skill encodes is the same judgment that goes into an architecture decision, and organizations should staff and reward it accordingly.

\subsection{Preserve a Comprehension Checkpoint Independent of Authorship}
\label{sec:impl-comprehension}

Convention-based comprehension means understanding a codebase because a shared skill shaped it into a familiar form rather than because you wrote it. This was a real and reliable route to comprehension in our sessions (\S\ref{sec:context-artifacts-gap}), and one the existing cognitive debt literature does not account for. But it only works when the convention is actually followed, which means tools need a way to check that it was. Surface which parts of a change conform to an existing convention file and which are genuinely novel, so review attention concentrates where convention cannot substitute for reading.

\subsection{Recognize AI-Assisted Development as a Trainable Competency}
\label{sec:impl-training}

Fluency with AI-assisted workflows behaved as a skill of its own in our data, separate from general engineering experience and not reliably predicted by seniority (\S\ref{sec:competence-paradox}). P8's company has already drawn the organizational conclusion, hiring and evaluating candidates partly on prompt quality. Training and onboarding elsewhere should follow the same logic rather than leave the competency to be picked up incidentally. Teach structured prompting, skill authoring, and verification discipline explicitly in onboarding, rather than assuming developers absorb them independently on the job.

\subsection{Better Skills: A Proof-of-Concept Instantiation}
\label{sec:better-skills}

Our observations make the case concrete. There is no standard way to author these files. Participants improvised PLAN.md, SKILL.md, AGENTS.md, and other instruction files to their own preferences, and developers at every level struggled to create or use high-quality skills without design guidance (\S\ref{sec:codified-convention}, \S\ref{sec:ai-skill-competence}). Authoring was iterative by necessity, and one participant edited their file mid-task after realizing they had forgotten to specify a return type. Reuse did not solve the problem either. Participants pulled public skills from sites such as skills.sh and getdesign.md, but many carry authoring smells \cite{Hong2026SkillSmells}; public skills go stale, and generic ones rarely encode the organization-specific conventions, naming, and libraries that make a skill useful. SP16's experience in \S\ref{sec:codified-convention} is the sharpest example. Adapting them costs the effort they were meant to save.

\begin{figure}[t]
    \centering
    \includegraphics[width=\columnwidth]{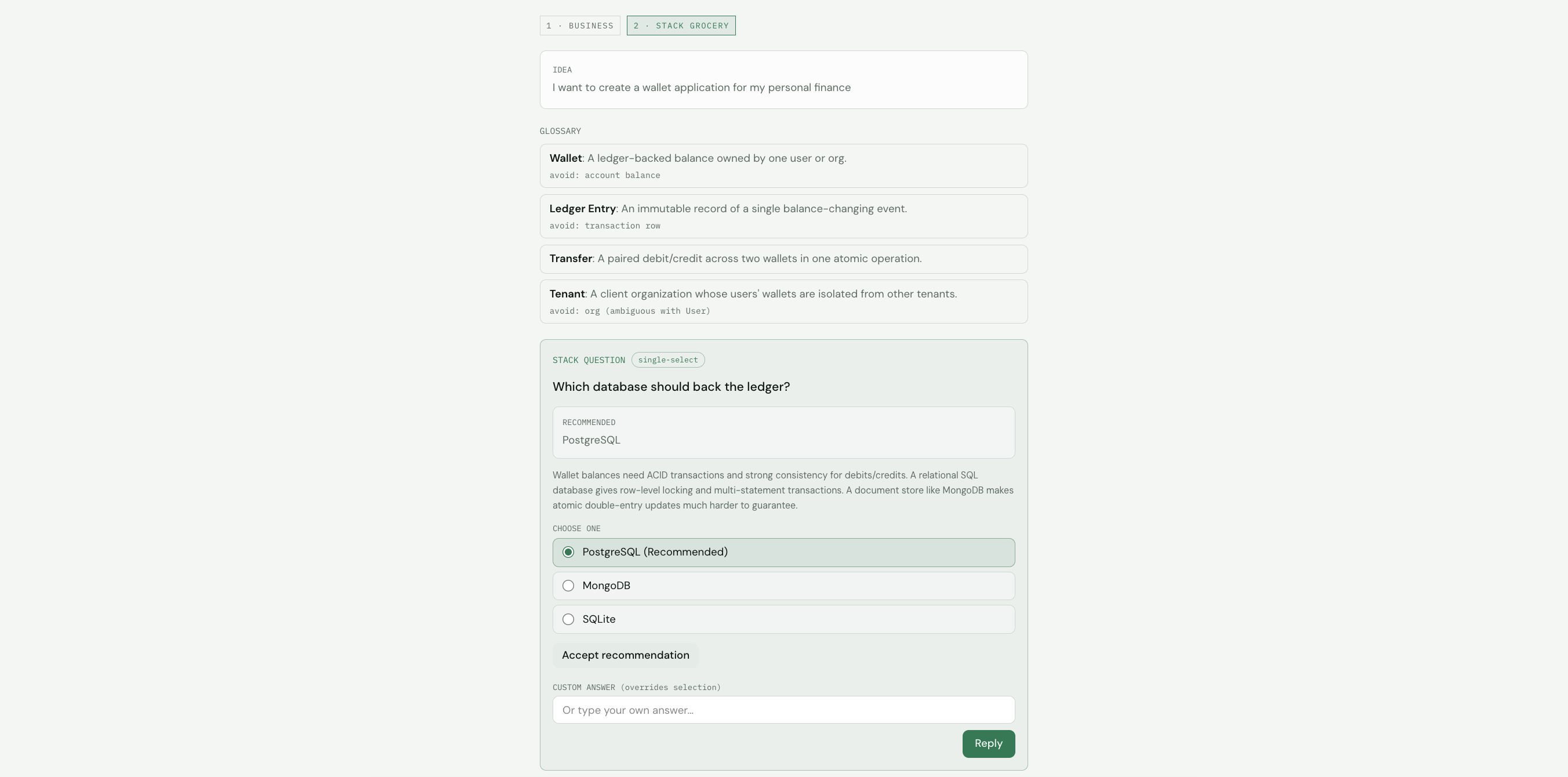}
    \caption{The Better Skills configuration workflow, which elicits project and stack choices while explaining recommendations and trade-offs.}
    \Description{A screenshot of step 2, Stack Grocery, in the Better Skills interface for a personal-finance wallet application. A glossary defines Wallet, Ledger Entry, Transfer, and Tenant. Below it, a single-select question asks which database should back the ledger. PostgreSQL is marked Recommended and selected, with text explaining that wallet balances need ACID transactions and strong consistency. MongoDB and SQLite appear as alternatives, followed by an Accept recommendation button, a custom-answer field, and a Reply button.}
    \label{fig:better-skills-builder}
\end{figure}

\begin{figure}[t]
    \centering
    \includegraphics[width=\columnwidth]{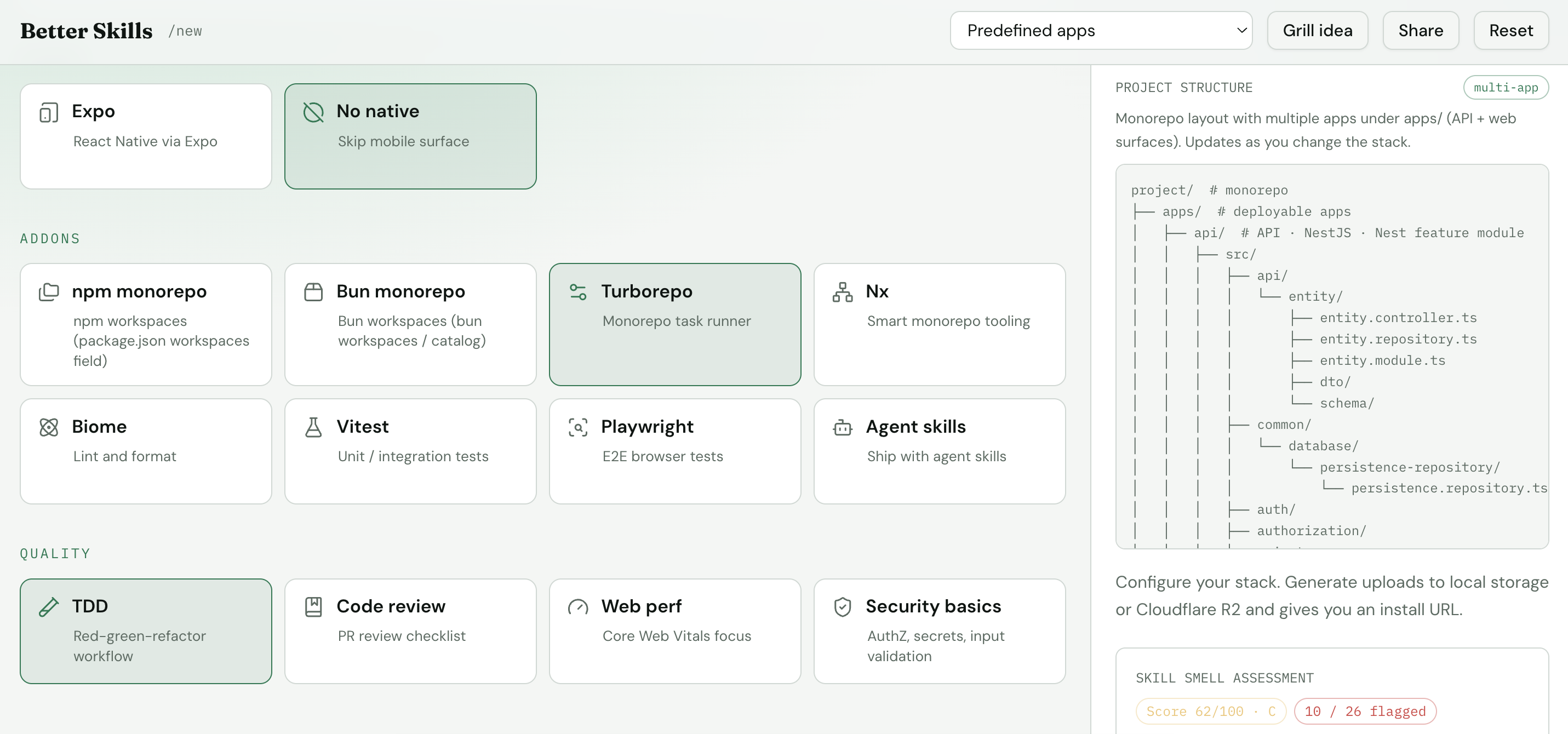}
    \caption{A Better Skills configuration and its generated project structure and quality assessment.}
    \Description{A wide screenshot of the Better Skills configuration interface. The left side shows selectable cards: Expo and selected No native; add-ons for npm monorepo, Bun monorepo, selected Turborepo, Nx, Biome, Vitest, Playwright, and Agent skills; and quality cards for selected test-driven development, Code review, Web perf, and Security basics. The right side previews a multi-app monorepo project tree with API and web applications and nested source, entity, database, authentication, and authorization folders. Beneath the tree, text explains that generation produces an install URL, and a Skill Smell Assessment reports a score of 62 out of 100, grade C, with 10 of 26 checks flagged.}
    \label{fig:better-skills-output}
\end{figure}

We built Better Skills in response. As this paper is findings-centric, Better Skills remains an early design probe rather than a system contribution. It is a working sketch of the instruction-quality implication above, supported only by informal feedback from three returning participants, and systematic evaluation remains future work (\S\ref{sec:limitations-and-future-work}). Rather than writing instruction files from scratch, users step through a guided interview called the Grill. It covers project shape, backend framework, API structure, code organization, authentication, database, and testing. A separate step-by-step builder renders the resulting folder structure live in a sidebar, allowing developers to review and adjust it. Better Skills generates skills from fixed templates rather than free-form model output, which keeps generation token-cheap. It also provides a CLI install command tied to a persistent URL teammates can reuse, along with example task prompts. A user can configure a backend project without an API surface, install the skill in one command, and immediately receive prompts for scaffolding, authentication, or database models (\autoref{fig:better-skills-builder}, \autoref{fig:better-skills-output}). Generated skills receive a smell score based on Hong et al.'s taxonomy \cite{Hong2026SkillSmells}, which developers can keep or remove. The template itself therefore steers authors away from the failures that dominate public skills.

Better Skills also provides decision support by recommending frameworks, architectural patterns, and tooling based on project requirements and scale. It discourages problematic pairings and explains each recommendation's rationale. Before generating a skill, a mandatory smell assessment checks the draft against Hong et al.'s 26-item taxonomy, returning a 0--100 score, letter grade, and plain-language verdict rather than a bare checklist. Generation cannot proceed until the assessment runs \footnote{Better Skills is available at \url{https://better-skills-web.vercel.app}. The prototype's Grill can be triggered by entering an idea containing the keyword ``wallet'' (e.g., `I want to make a wallet application for my personal finance''), after which it presents the corresponding business-validation questions. To enable no-cost public access without an API key, the prototype substitutes scripted, realistic responses for the interview and smell assessment; this mode exists only for accessibility, and none of its output is used as evidence in this paper.}. The design intent is for this guidance to double as instruction, helping beginners understand architectural choices and quality feedback in context. This addresses the fifth implication above through the artifact addressing the third.

As a first check, we walked three contextual inquiry participants (P2, P7, P8) through the prototype in short follow-up interviews (\S\ref{sec:prototype-development}). P2 said the per-step explanations made the process legible, because he understood what was happening at each choice rather than watching configuration happen to him. P7 and P8 liked assembling a clean skill step by step but wanted more architectural and coding patterns covered. We read this as encouraging for legibility and smell-free structure, and as a signal that the configuration space is too shallow.

\section{Limitations and Future Work} \label{sec:limitations-and-future-work}
Our sample carries the biases of its recruitment. The survey was seeded in our professional networks and spread by snowball referral and public Facebook developer communities, and contextual inquiry participants were a convenience sample from our own network (\S\ref{sec:participants}). Both routes favor developers already comfortable enough with AI tooling to volunteer for a study about it, within a single country. Developers who have abandoned or never adopted these tools are underrepresented or absent. No substantially new themes emerged after eleven sessions, and three confirmatory sessions were added (\S\ref{sec:participants}), which gives reasonable confidence in the stability of the seven themes for this population. Even so, all fourteen sessions involved one class of application. The specification-labor, delegation-boundary, and verification patterns we report may look different on a large legacy codebase, in a regulated domain such as finance or healthcare, or on a task with a harder ground truth than a management-system web app.

These boundaries set the most direct agenda for extension. A comparative study across regions with different cost structures, tooling access, seniority distributions, and English proficiency (\S\ref{sec:competence-paradox}) would separate what is specific to a cost-constrained, fast-growing market from what generalizes. We suspect the token-economy behaviors of \S\ref{sec:token-economy} are sharper here than where AI subscriptions are a smaller fraction of income. Extending the task to legacy codebases, regulated domains, and non-web platforms would test whether the delegation boundaries of \S\ref{sec:retained-half} hold beyond the task type we studied.

The quantitative strand is cross-sectional self-report, and its claims are modest by construction. Single items carry their own measurement noise, self-report associations can be inflated by common-method variance, and a cross-sectional design cannot order relationships temporally. The correlations we highlight, such as careful reading tracking fewer accepted-then-buggy episodes, are consistent with the mechanism our qualitative data describe but cannot establish it causally. The 45--60 minute session window likewise cannot speak to the timescale at which cognitive debt is usually studied. Our ownership probe measures immediate recall right after a task, not what a developer can still explain weeks later. A longitudinal design would address both gaps. It would test whether shared-convention comprehension (\S\ref{sec:ownership-comprehension}) holds up over months or erodes on the timescale cognitive-debt research documents for individually authored understanding \cite{Kosmyna2025CognitiveDebt}. It would also pair observed practices with repository-level outcomes such as commit size, review latency, and defect rate, so our design implications (\S\ref{sec:design-implications-recommendations}) can be evaluated against project outcomes rather than self-report alone.

Our findings on persistent instruction files rest on what participants told us about their skills and rules files, not on a direct audit. We did not systematically collect and score participants' own SKILL.md or CLAUDE.md files. The connection we draw in \S\ref{sec:context-artifacts-gap} between our qualitative pattern and the skill-smell prevalence Hong et al. report \cite{Hong2026SkillSmells} is therefore a convergence across two independent studies, not a within-study measurement. Closing this gap is the most direct extension of this work. Auditing participants' own files against a known smell taxonomy \cite{Hong2026SkillSmells}, for instance through Better Skills's analysis component (\S\ref{sec:better-skills}), would test the link between skill quality and outcomes such as SP16's reported velocity gain within a single sample.

Better Skills itself shares this limitation, and more. We present it as an early design probe rather than a system contribution, and our evidence for it is informal feedback interviews with three returning participants (\S\ref{sec:better-skills}). That feedback is encouraging but no substitute for evaluation. None of the benefits we designed it for have been systematically tested, including fewer authoring smells, lower token cost, faster onboarding, and learning effects for beginners. In particular, we have not run its own smell-detection component over its generated skills and a sample of public ones to substantiate the ``smell-free by design'' claim, which remains a design intent rather than a measured result. The current implementation also only supports JavaScript and TypeScript stacks. The template architecture is language-agnostic by design, but we have not built generators for other ecosystems. A deployment study is the natural companion to the audit above, testing whether skills from its guided workflow outperform hand-written or public ones, extended to non-JavaScript stacks and validated with a real smell-count comparison.

Our findings also point to organizational questions we lacked the data to answer. Who authors and maintains shared skills, how is that work recognized, and is hiring on prompt quality, as P8's company already does, becoming common? Finally, these tools change quickly, and our data capture a specific window in that trajectory.

\section{Conclusion} \label{sec:conclusion}
We set out to find where human effort goes once an AI agent can produce working code on request. Across a 116-respondent survey and 14 observed development sessions with paired interviews, the answer converges on a single shape: effort does not disappear; it relocates. It moves into planning before a prompt is sent, deciding what to hand over and what to keep, checking output closely enough to catch failures a passing demo will not reveal, and building the shared conventions that keep a team's AI output legible to itself. The developers who benefit most are not those who trust the tool the most, but those who have rebuilt, often without naming it, the ordinary discipline of software engineering around a conversational interface that does not supply that discipline on its own. We call this the competence paradox of vibe coding: the central design question is not simply how to make AI more capable, but how to make the human disciplines it still depends on—specification, verification, and codified convention—easier to build, share, and trust. That last discipline remains the least supported. Persistent instruction files were the practice our participants converged on to make vibe coding sustainable at a team level, and the same practice that failed hardest when done badly. Better Skills, the early design probe we describe in \S\ref{sec:better-skills}, sketches one answer by treating shared AI instructions as artifacts deserving the same authoring care, review, and tooling as the code they govern. Whether it delivers remains for the work outlined in \S\ref{sec:limitations-and-future-work}; if vibe coding is to mature into engineering rather than remain a name for improvisation, the scaffolding developers are already quietly building by hand must become something the tools help them build well.

\bibliographystyle{ACM-Reference-Format}
\bibliography{references}

@online{Karpathy2025VibeCoding,
  author  = {Andrej Karpathy},
  title   = {There's a new kind of coding I call ``vibe coding''},
  year    = {2025},
  month   = feb,
  day     = {2},
  url     = {https://x.com/karpathy/status/1886192184808149383},
  note    = {Post on X (formerly Twitter). Accessed: 2026-07-03}
}

@article{Saunders2018Saturation,
  author  = {Saunders, Benjamin and Sim, Julius and Kingstone, Tom and Baker, Shula and Waterfield, Jackie and Bartlam, Bernadette and Burroughs, Heather and Jinks, Clare},
  title   = {Saturation in qualitative research: exploring its conceptualization and operationalization},
  journal = {Quality \& Quantity},
  year    = {2018},
  volume  = {52},
  number  = {4},
  pages   = {1893--1907},
  doi     = {10.1007/s11135-017-0574-8},
  url     = {https://doi.org/10.1007/s11135-017-0574-8}
}

@incollection{Courage2005Field,
  author    = {Courage, Catherine and Baxter, Kathy},
  title     = {Field Studies},
  booktitle = {Understanding Your Users: A Practical Guide to User Requirements Methods, Tools, and Techniques},
  publisher = {Morgan Kaufmann},
  address   = {San Francisco},
  year      = {2005},
  pages     = {562--633},
  series    = {Interactive Technologies},
  isbn      = {978-1-55860-935-8},
  doi       = {10.1016/B978-155860935-8/50043-9}
}

@article{Ray2025VibeCodingReview,
  author = {Partha Pratim Ray},
  title = {A Review on Vibe Coding: Fundamentals, State-of-the-art, Challenges and Future Directions},
  journal = {TechRxiv},
  volume = {2025},
  number = {0509},
  pages = {},
  year = {2025},
  doi = {10.36227/techrxiv.174681482.27435614/v1},
  URL = {https://www.techrxiv.org/doi/abs/10.36227/techrxiv.174681482.27435614/v1},
  eprint = {https://www.techrxiv.org/doi/pdf/10.36227/techrxiv.174681482.27435614/v1},
}

@misc{Drosos2025VibeCoding,
      title={Vibe coding: programming through conversation with artificial intelligence}, 
      author={Advait Sarkar and Ian Drosos},
      year={2025},
      eprint={2506.23253},
      archivePrefix={arXiv},
      primaryClass={cs.HC},
      url={https://doi.org/10.48550/arXiv.2506.23253},
      doi={10.48550/arXiv.2506.23253}
}

@article{Campbell2013Coding,
 author = {Campbell, John and Quincy, Charles and Osserman, Jordan and Pedersen, Ove},
 year = {2013},
 month = {08},
 pages = {294-320},
 title = {Coding In-depth Semistructured Interviews: Problems of Unitization and Intercoder Reliability and Agreement},
 volume = {42},
 journal = {Sociological Methods \& Research},
 doi = {10.1177/0049124113500475}
}

@article{Horvat2025WhatIsVibe,
    author = {Marko Horvat },
    title = {What is Vibe coding and when should you use it (or not)?},
    journal = {TechRxiv},
    volume = {2025},
    number = {0807},
    pages = {},
    year = {2025},
    doi = {10.36227/techrxiv.175459780.03758839/v1},
    URL = {https://www.techrxiv.org/doi/abs/10.36227/techrxiv.175459780.03758839/v1},
}

@misc{Sapkota2025VibeAgentic,
      title={Vibe Coding vs. Agentic Coding: Fundamentals and Practical Implications of Agentic AI}, 
      author={Ranjan Sapkota and Konstantinos I. Roumeliotis and Manoj Karkee},
      year={2025},
      eprint={2505.19443},
      archivePrefix={arXiv},
      primaryClass={cs.SE},
      url={https://doi.org/10.48550/arXiv.2505.19443},
      doi={10.48550/arXiv.2505.19443}
}

@inproceedings{Li2026VibeCoding,
    author = {Li, Jie and Hou, Youyang and Lin, Laura and Zhu, Ruihao and Cao, Hancheng and El Ali, Abdallah},
    title = {Vibe Coding in Product Teams: Reconfiguring AI-Assisted Workflows, Prototyping, and Collaboration},
    year = {2026},
    isbn = {9798400724299},
    publisher = {Association for Computing Machinery},
    address = {New York, NY, USA},
    url = {https://doi.org/10.1145/3808045.3808062},
    doi = {10.1145/3808045.3808062},
    booktitle = {Proceedings of the 5th Annual Symposium on Human-Computer Interaction for Work},
    articleno = {1},
    numpages = {16},
    series = {CHIWORK '26}
}

@inproceedings{Subramonyam2025Prototyping,
    author = {Subramonyam, Hari and Thakkar, Divy and Ku, Andrew and Dieber, Juergen and Sinha, Anoop K.},
    title = {Prototyping with Prompts: Emerging Approaches and Challenges in Generative AI Design for Collaborative Software Teams},
    year = {2025},
    isbn = {9798400713941},
    publisher = {Association for Computing Machinery},
    address = {New York, NY, USA},
    url = {https://doi.org/10.1145/3706598.3713166},
    doi = {10.1145/3706598.3713166},
    booktitle = {Proceedings of the 2025 CHI Conference on Human Factors in Computing Systems},
    articleno = {882},
    numpages = {22},
    series = {CHI '25}
}

@inproceedings{Li2025From,
    author = {Li, Zhuoyan and Zhu, Hangxiao and Lu, Zhuoran and Xiao, Ziang and Yin, Ming},
    title = {From Text to Trust: Empowering AI-assisted Decision Making with Adaptive LLM-powered Analysis},
    year = {2025},
    isbn = {9798400713941},
    publisher = {Association for Computing Machinery},
    address = {New York, NY, USA},
    url = {https://doi.org/10.1145/3706598.3713133},
    doi = {10.1145/3706598.3713133},
    booktitle = {Proceedings of the 2025 CHI Conference on Human Factors in Computing Systems},
    articleno = {1193},
    numpages = {18},
    series = {CHI '25}
}

@book{Ericsson1984Protocol,
  author    = {K. Anders Ericsson and Herbert A. Simon},
  title     = {Protocol Analysis: Verbal Reports as Data},
  year      = {1984},
  publisher = {MIT Press},
  address   = {Cambridge, MA},
  isbn      = {9780262050296}
}

@inproceedings{Mozannar2024Reading,
    author = {Mozannar, Hussein and Bansal, Gagan and Fourney, Adam and Horvitz, Eric},
    title = {Reading Between the Lines: Modeling User Behavior and Costs in AI-Assisted Programming},
    year = {2024},
    isbn = {9798400703300},
    publisher = {Association for Computing Machinery},
    address = {New York, NY, USA},
    url = {https://doi.org/10.1145/3613904.3641936},
    doi = {10.1145/3613904.3641936},
    booktitle = {Proceedings of the 2024 CHI Conference on Human Factors in Computing Systems},
    articleno = {142},
    numpages = {16},
    location = {Honolulu, HI, USA},
    series = {CHI '24}
}

@inproceedings{Zamfirescu2025Beyond,
    author = {Zamfirescu-Pereira, J.D. and Jun, Eunice and Terry, Michael and Yang, Qian and Hartmann, Bjoern},
    title = {Beyond Code Generation: LLM-supported Exploration of the Program Design Space},
    year = {2025},
    isbn = {9798400713941},
    publisher = {Association for Computing Machinery},
    address = {New York, NY, USA},
    url = {https://doi.org/10.1145/3706598.3714154},
    doi = {10.1145/3706598.3714154},
    booktitle = {Proceedings of the 2025 CHI Conference on Human Factors in Computing Systems},
    articleno = {153},
    numpages = {17},
    series = {CHI '25}
}

@article{Braun2019Reflexive,
  author  = {Virginia Braun and Victoria Clarke},
  title   = {Reflecting on Reflexive Thematic Analysis},
  journal = {Qualitative Research in Sport, Exercise and Health},
  volume  = {11},
  number  = {4},
  pages   = {589--597},
  year    = {2019},
  doi     = {10.1080/2159676X.2019.1628806},
  url     = {https://doi.org/10.1080/2159676X.2019.1628806}
}

@misc{Kosmyna2025CognitiveDebt,
      title={Your Brain on ChatGPT: Accumulation of Cognitive Debt when Using an AI Assistant for Essay Writing Task}, 
      author={Nataliya Kosmyna and Eugene Hauptmann and Ye Tong Yuan and Jessica Situ and Xian-Hao Liao and Ashly Vivian Beresnitzky and Iris Braunstein and Pattie Maes},
      year={2025},
      eprint={2506.08872},
      archivePrefix={arXiv},
      primaryClass={cs.AI},
      url={https://doi.org/10.48550/arXiv.2506.08872},
      doi={10.48550/arXiv.2506.08872}
}

@inproceedings{Earle2025DreamGarden,
    author = {Earle, Sam and Parajuli, Samyak and Banburski-Fahey, Andrzej},
    title = {DreamGarden: A Designer Assistant for Growing Games from a Single Prompt},
    year = {2025},
    isbn = {9798400713941},
    publisher = {Association for Computing Machinery},
    address = {New York, NY, USA},
    url = {https://doi.org/10.1145/3706598.3714233},
    doi = {10.1145/3706598.3714233},
    booktitle = {Proceedings of the 2025 CHI Conference on Human Factors in Computing Systems},
    articleno = {57},
    numpages = {19},
    series = {CHI '25}
}

@inproceedings{Yan2024Ivie,
    author = {Yan, Litao and Hwang, Alyssa and Wu, Zhiyuan and Head, Andrew},
    title = {Ivie: Lightweight Anchored Explanations of Just-Generated Code},
    year = {2024},
    isbn = {9798400703300},
    publisher = {Association for Computing Machinery},
    address = {New York, NY, USA},
    url = {https://doi.org/10.1145/3613904.3642239},
    doi = {10.1145/3613904.3642239},
    booktitle = {Proceedings of the 2024 CHI Conference on Human Factors in Computing Systems},
    articleno = {140},
    numpages = {15},
    location = {Honolulu, HI, USA},
    series = {CHI '24}
}

@inproceedings{Gebreegziabher2025Supporting,
    author = {Gebreegziabher, Simret Araya and Yang, Yukun and Glassman, Elena L. and Li, Toby Jia-Jun},
    title = {Supporting Co-Adaptive Machine Teaching through Human Concept Learning and Cognitive Theories},
    year = {2025},
    isbn = {9798400713941},
    publisher = {Association for Computing Machinery},
    address = {New York, NY, USA},
    url = {https://doi.org/10.1145/3706598.3713708},
    doi = {10.1145/3706598.3713708},
    booktitle = {Proceedings of the 2025 CHI Conference on Human Factors in Computing Systems},
    articleno = {533},
    numpages = {18},
    series = {CHI '25}
}

@book{likert1932technique,
  title={A Technique for the Measurement of Attitudes},
  author={Likert, Rensis},
  number={nos. 136-165},
  isbn={9780598579874},
  lccn={33012634},
  series={A Technique for the Measurement of Attitudes},
  year={1932},
  publisher={Columbia University},
  address={New York, NY, USA}
}

@book{mackenzie2013human,
  author    = {MacKenzie, I. Scott},
  title     = {Human-Computer Interaction: An Empirical Research Perspective},
  year     = {2013},
  address  = {San Francisco, CA, United States},
  publisher = {Morgan Kaufmann Publishers Inc.},
  isbn      = {978-0-12-405865-1},
  pagetotal = {370}
}

@inproceedings{Vaithilingam2022Expectation,
    author = {Vaithilingam, Priyan and Zhang, Tianyi and Glassman, Elena L.},
    title = {Expectation vs. Experience: Evaluating the Usability of Code Generation Tools Powered by Large Language Models},
    year = {2022},
    isbn = {9781450391566},
    publisher = {Association for Computing Machinery},
    address = {New York, NY, USA},
    url = {https://doi.org/10.1145/3491101.3519665},
    doi = {10.1145/3491101.3519665},
    booktitle = {Extended Abstracts of the 2022 CHI Conference on Human Factors in Computing Systems},
    articleno = {332},
    numpages = {7},
    location = {New Orleans, LA, USA},
    series = {CHI EA '22}
}

@inproceedings{Ferdowsi2024Validating,
    author = {Ferdowsi, Kasra and Huang, Ruanqianqian (Lisa) and James, Michael B. and Polikarpova, Nadia and Lerner, Sorin},
    title = {Validating AI-Generated Code with Live Programming},
    year = {2024},
    isbn = {9798400703300},
    publisher = {Association for Computing Machinery},
    address = {New York, NY, USA},
    url = {https://doi.org/10.1145/3613904.3642495},
    doi = {10.1145/3613904.3642495},
    booktitle = {Proceedings of the 2024 CHI Conference on Human Factors in Computing Systems},
    articleno = {143},
    numpages = {8},
    location = {Honolulu, HI, USA},
    series = {CHI '24}
}

@inproceedings{Wang2024UML,
  author    = {Wang, Beian and Wang, Chong and Liang, Peng and Li, Bing and Zeng, Cheng},
  booktitle = {2024 IEEE International Conference on Software Services Engineering (SSE)},
  title     = {How LLMs Aid in UML Modeling: An Exploratory Study with Novice Analysts},
  year      = {2024},
  pages     = {249--257},
  publisher = {IEEE},
  address   = {Piscataway, NJ},
  doi       = {10.1109/SSE62657.2024.00046}
}

@inproceedings{Epperson2025Interactive,
    author = {Epperson, Will and Bansal, Gagan and Dibia, Victor C and Fourney, Adam and Gerrits, Jack and Zhu, Erkang (Eric) and Amershi, Saleema},
    title = {Interactive Debugging and Steering of Multi-Agent AI Systems},
    year = {2025},
    isbn = {9798400713941},
    publisher = {Association for Computing Machinery},
    address = {New York, NY, USA},
    url = {https://doi.org/10.1145/3706598.3713581},
    doi = {10.1145/3706598.3713581},
    booktitle = {Proceedings of the 2025 CHI Conference on Human Factors in Computing Systems},
    articleno = {156},
    numpages = {15},
    series = {CHI '25}
}

@article{Goddard2012AutomationBias,
  author  = {Kate Goddard and Abdul Roudsari and Jeremy C. Wyatt},
  title   = {Automation Bias: A Systematic Review of Frequency, Effect Mediators, and Mitigators},
  journal = {Journal of the American Medical Informatics Association},
  volume  = {19},
  number  = {1},
  pages   = {121--127},
  year    = {2012},
  month   = jan,
  doi     = {10.1136/amiajnl-2011-000089},
  pmid    = {21685142},
  pmcid   = {PMC3240751},
  url      = {https://doi.org/10.1136/amiajnl-2011-000089}
}

@inproceedings{Khurana2025DoItForMe,
    author = {Khurana, Anjali and Su, Xiaotian and Wang, April Yi and Chilana, Parmit K},
    title = {Do It For Me vs. Do It With Me: Investigating User Perceptions of Different Paradigms of Automation in Copilots for Feature-Rich Software},
    year = {2025},
    isbn = {9798400713941},
    publisher = {Association for Computing Machinery},
    address = {New York, NY, USA},
    url = {https://doi.org/10.1145/3706598.3713431},
    doi = {10.1145/3706598.3713431},
    booktitle = {Proceedings of the 2025 CHI Conference on Human Factors in Computing Systems},
    articleno = {880},
    numpages = {18},
    series = {CHI '25}
}

@inproceedings{Tholander2026VibeCoding,
    author = {Tholander, Jakob and Jonsson, Martin},
    title = {Vibe Coding Entanglements – Repositioning Boundaries of Intention, Authorship, and Responsibility in Programming with Generative AI},
    year = {2026},
    isbn = {9798400722783},
    publisher = {Association for Computing Machinery},
    address = {New York, NY, USA},
    url = {https://doi.org/10.1145/3772318.3791847},
    doi = {10.1145/3772318.3791847},
    booktitle = {Proceedings of the 2026 CHI Conference on Human Factors in Computing Systems},
    articleno = {90},
    numpages = {16},
    series = {CHI '26}
}

@inproceedings{Feng2026AgencyAI,
    author = {Feng, Dana and Yun, Bhada and Yi Wang, April},
    title = {From Junior to Senior: Allocating Agency and Navigating Professional Growth in Agentic AI-Mediated Software Engineering},
    year = {2026},
    isbn = {9798400722783},
    publisher = {Association for Computing Machinery},
    address = {New York, NY, USA},
    url = {https://doi.org/10.1145/3772318.3791642},
    doi = {10.1145/3772318.3791642},
    booktitle = {Proceedings of the 2026 CHI Conference on Human Factors in Computing Systems},
    articleno = {596},
    numpages = {24},
    series = {CHI '26}
}

@inproceedings{Yen2025CodeShaping,
    author = {Yen, Ryan and Zhao, Jian and Vogel, Daniel},
    title = {Code Shaping: Iterative Code Editing with Free-form AI-Interpreted Sketching},
    year = {2025},
    isbn = {9798400713941},
    publisher = {Association for Computing Machinery},
    address = {New York, NY, USA},
    url = {https://doi.org/10.1145/3706598.3713822},
    doi = {10.1145/3706598.3713822},
    booktitle = {Proceedings of the 2025 CHI Conference on Human Factors in Computing Systems},
    articleno = {872},
    numpages = {17},
    series = {CHI '25}
}

@inproceedings{He2025Prompting,
    author = {He, Zeyu and Naphade, Saniya and Huang, Ting-Hao Kenneth},
    title = {Prompting in the Dark: Assessing Human Performance in Prompt Engineering for Data Labeling When Gold Labels Are Absent},
    year = {2025},
    isbn = {9798400713941},
    publisher = {Association for Computing Machinery},
    address = {New York, NY, USA},
    url = {https://doi.org/10.1145/3706598.3714319},
    doi = {10.1145/3706598.3714319},
    booktitle = {Proceedings of the 2025 CHI Conference on Human Factors in Computing Systems},
    articleno = {1195},
    numpages = {33},
    series = {CHI '25}
}

@article{Ziegler2024Copilot,
    author = {Ziegler, Albert and Kalliamvakou, Eirini and Li, X. Alice and Rice, Andrew and Rifkin, Devon and Simister, Shawn and Sittampalam, Ganesh and Aftandilian, Edward},
    title = {Measuring GitHub Copilot's Impact on Productivity},
    year = {2024},
    issue_date = {March 2024},
    publisher = {Association for Computing Machinery},
    address = {New York, NY, USA},
    volume = {67},
    number = {3},
    issn = {0001-0782},
    url = {https://doi.org/10.1145/3633453},
    doi = {10.1145/3633453},
    journal = {Commun. ACM},
    month = feb,
    pages = {54–63},
    numpages = {10}
}

@misc{Voronin2024Development,
     title={Development and Evaluation of an LLM-Based Tool for Automatically Building Web Applications},
     url={https://hdl.handle.net/1721.1/156803},
     school={Massachusetts Institute of Technology},
     author={Voronin, Diana Nguyen},
     year={2024},
     month=May
 }

@inproceedings{Pu2025Assistance,
    author = {Pu, Kevin and Lazaro, Daniel and Arawjo, Ian and Xia, Haijun and Xiao, Ziang and Grossman, Tovi and Chen, Yan},
    title = {Assistance or Disruption? Exploring and Evaluating the Design and Trade-offs of Proactive AI Programming Support},
    year = {2025},
    isbn = {9798400713941},
    publisher = {Association for Computing Machinery},
    address = {New York, NY, USA},
    url = {https://doi.org/10.1145/3706598.3713357},
    doi = {10.1145/3706598.3713357},
    booktitle = {Proceedings of the 2025 CHI Conference on Human Factors in Computing Systems},
    articleno = {152},
    numpages = {21},
    series = {CHI '25}
}

@inproceedings{OBrien2024Todo,
    author = {OBrien, David and Biswas, Sumon and Imtiaz, Sayem Mohammad and Abdalkareem, Rabe and Shihab, Emad and Rajan, Hridesh},
    title = {Are Prompt Engineering and TODO Comments Friends or Foes? An Evaluation on GitHub Copilot},
    year = {2024},
    isbn = {9798400702174},
    publisher = {Association for Computing Machinery},
    address = {New York, NY, USA},
    url = {https://doi.org/10.1145/3597503.3639176},
    doi = {10.1145/3597503.3639176},
    booktitle = {Proceedings of the IEEE/ACM 46th International Conference on Software Engineering},
    articleno = {219},
    numpages = {13},
    location = {Lisbon, Portugal},
    series = {ICSE '24}
}

@article{Barke2023Grounded,
    author = {Barke, Shraddha and James, Michael B. and Polikarpova, Nadia},
    title = {Grounded Copilot: How Programmers Interact with Code-Generating Models},
    year = {2023},
    issue_date = {April 2023},
    publisher = {Association for Computing Machinery},
    address = {New York, NY, USA},
    volume = {7},
    number = {OOPSLA1},
    url = {https://doi.org/10.1145/3586030},
    doi = {10.1145/3586030},
    journal = {Proc. ACM Program. Lang.},
    month = apr,
    articleno = {78},
    numpages = {27}
}

@inproceedings{Pickering2025HowHumans,
    author = {Pickering, Madison and Williams, Helena and Gan, Alison and He, Weijia and Park, Hyojae and Piedrahita Velez, Francisco and Littman, Michael L. and Ur, Blase},
    title = {How Humans Communicate Programming Tasks in Natural Language and Implications For End-User Programming with LLMs},
    year = {2025},
    isbn = {9798400713941},
    publisher = {Association for Computing Machinery},
    address = {New York, NY, USA},
    url = {https://doi.org/10.1145/3706598.3713271},
    doi = {10.1145/3706598.3713271},
    booktitle = {Proceedings of the 2025 CHI Conference on Human Factors in Computing Systems},
    articleno = {875},
    numpages = {34},
    series = {CHI '25}
}

@inproceedings{Zamfirescu2023WhyJohnny,
    author = {Zamfirescu-Pereira, J.D. and Wong, Richmond Y. and Hartmann, Bjoern and Yang, Qian},
    title = {Why Johnny Can’t Prompt: How Non-AI Experts Try (and Fail) to Design LLM Prompts},
    year = {2023},
    isbn = {9781450394215},
    publisher = {Association for Computing Machinery},
    address = {New York, NY, USA},
    url = {https://doi.org/10.1145/3544548.3581388},
    doi = {10.1145/3544548.3581388},
    booktitle = {Proceedings of the 2023 CHI Conference on Human Factors in Computing Systems},
    articleno = {437},
    numpages = {21},
    location = {Hamburg, Germany},
    series = {CHI '23}
}

@inproceedings{Chen2025NeedHelp,
    author = {Chen, Valerie and Zhu, Alan and Zhao, Sebastian and Mozannar, Hussein and Sontag, David and Talwalkar, Ameet},
    title = {Need Help? Designing Proactive AI Assistants for Programming},
    year = {2025},
    isbn = {9798400713941},
    publisher = {Association for Computing Machinery},
    address = {New York, NY, USA},
    url = {https://doi.org/10.1145/3706598.3714002},
    doi = {10.1145/3706598.3714002},
    booktitle = {Proceedings of the 2025 CHI Conference on Human Factors in Computing Systems},
    articleno = {881},
    numpages = {18},
    series = {CHI '25}
}

@inproceedings{Subramonyam2024Bridging,
    author = {Subramonyam, Hari and Pea, Roy and Pondoc, Christopher and Agrawala, Maneesh and Seifert, Colleen},
    title = {Bridging the Gulf of Envisioning: Cognitive Challenges in Prompt Based Interactions with LLMs},
    year = {2024},
    isbn = {9798400703300},
    publisher = {Association for Computing Machinery},
    address = {New York, NY, USA},
    url = {https://doi.org/10.1145/3613904.3642754},
    doi = {10.1145/3613904.3642754},
    booktitle = {Proceedings of the 2024 CHI Conference on Human Factors in Computing Systems},
    articleno = {1039},
    numpages = {19},
    location = {Honolulu, HI, USA},
    series = {CHI '24}
}

@inproceedings{Wong2025ItIsASpectrum,
    author = {Wong, Novia and Cheng, Nai-Yu and Oewel, Bruna and Genuario, Katherine E and Stoeckl, SarahElizabeth and Schueller, Stephen M. and Ahmed, Iftekhar and van der Hoek, Andr\'{e} and Reddy, Madhu},
    title = { 'It's a spectrum': Exploring Autonomy, Competence, and Relatedness in Software Development Processes and Tools},
    year = {2025},
    isbn = {9798400713941},
    publisher = {Association for Computing Machinery},
    address = {New York, NY, USA},
    url = {https://doi.org/10.1145/3706598.3713250},
    doi = {10.1145/3706598.3713250},
    booktitle = {Proceedings of the 2025 CHI Conference on Human Factors in Computing Systems},
    articleno = {151},
    numpages = {19},
    series = {CHI '25}
}

@inproceedings{Schecter2025Role,
    author = {Schecter, Aaron and Richardson, Benjamin},
    title = {How the Role of Generative AI Shapes Perceptions of Value in Human-AI Collaborative Work},
    year = {2025},
    isbn = {9798400713941},
    publisher = {Association for Computing Machinery},
    address = {New York, NY, USA},
    url = {https://doi.org/10.1145/3706598.3713946},
    doi = {10.1145/3706598.3713946},
    booktitle = {Proceedings of the 2025 CHI Conference on Human Factors in Computing Systems},
    articleno = {530},
    numpages = {15},
    series = {CHI '25}
}

@inproceedings{Zheng2025NonAI,
    author = {Zheng, Qingxiao and Chen, Minrui and Park, Hyanghee and Xu, Zhongwei and Huang, Yun},
    title = {Evaluating Non-AI Experts' Interaction with AI: A Case Study In Library Context},
    year = {2025},
    isbn = {9798400713941},
    publisher = {Association for Computing Machinery},
    address = {New York, NY, USA},
    url = {https://doi.org/10.1145/3706598.3714219},
    doi = {10.1145/3706598.3714219},
    booktitle = {Proceedings of the 2025 CHI Conference on Human Factors in Computing Systems},
    articleno = {1192},
    numpages = {20},
    series = {CHI '25}
}

@inproceedings{Ali2025EvenThough,
    author = {Ali, Murtaza and Dasgupta, Sayamindu},
    title = {"Even Though I Went Through Everything, I Didn't Feel Like I Learned a Lot": Insights From Experiences of Non-Computer Science Students Learning to Code},
    year = {2025},
    isbn = {9798400713941},
    publisher = {Association for Computing Machinery},
    address = {New York, NY, USA},
    url = {https://doi.org/10.1145/3706598.3713624},
    doi = {10.1145/3706598.3713624},
    booktitle = {Proceedings of the 2025 CHI Conference on Human Factors in Computing Systems},
    articleno = {49},
    numpages = {18},
    series = {CHI '25}
}

@inproceedings{Bunch2026Trustworthiness,
    author = {Bunch, Olivia and Taeb, Maryam},
    title = {Trustworthiness of Vibe Coding from the NIST AI RMF Perspective},
    year = {2026},
    isbn = {9798400720628},
    publisher = {Association for Computing Machinery},
    address = {New York, NY, USA},
    url = {https://doi.org/10.1145/3746467.3801530},
    doi = {10.1145/3746467.3801530},
    booktitle = {Proceedings of the 2026 ACM Southeast Conference},
    pages = {229–234},
    numpages = {6},
    location = {Troy University, Troy, AL, USA},
    series = {ACMSE 2026}
}

@inproceedings{Geng2025Exploring,
    author = {Geng, Francis and Shah, Anshul and Li, Haolin and Mulla, Nawab and Swanson, Steve and Soosai Raj, Gerald and Zingaro, Daniel and Porter, Leo},
    title = {Exploring Student-AI Interactions in Vibe Coding},
    year = {2026},
    isbn = {9798400723520},
    publisher = {Association for Computing Machinery},
    address = {New York, NY, USA},
    url = {https://doi.org/10.1145/3786228.3786236},
    doi = {10.1145/3786228.3786236},
    booktitle = {Proceedings of the 28th Australasian Computing Education Conference},
    pages = {45–54},
    numpages = {10},
    series = {ACE '26}
}

@inproceedings{Mowar2025CodeA11y,
    author = {Mowar, Peya and Peng, Yi-Hao and Wu, Jason and Steinfeld, Aaron and Bigham, Jeffrey P},
    title = {CodeA11y: Making AI Coding Assistants Useful for Accessible Web Development},
    year = {2025},
    isbn = {9798400713941},
    publisher = {Association for Computing Machinery},
    address = {New York, NY, USA},
    url = {https://doi.org/10.1145/3706598.3713335},
    doi = {10.1145/3706598.3713335},
    booktitle = {Proceedings of the 2025 CHI Conference on Human Factors in Computing Systems},
    articleno = {45},
    numpages = {15},
    series = {CHI '25}
}

@misc{openai_chatgpt,
  author       = {{OpenAI}},
  title        = {ChatGPT},
  year         = {2026},
  howpublished = {\url{https://chatgpt.com/}},
  note         = {Large language model-based chat platform}
}

@misc{anthropic_claude,
  author       = {{Anthropic}},
  title        = {Claude},
  year         = {2026},
  howpublished = {\url{https://claude.ai/}},
  note         = {Large language model-based chat platform}
}

@misc{github_copilot,
  author       = {{GitHub}},
  title        = {GitHub Copilot},
  year         = {2026},
  howpublished = {\url{https://github.com/features/copilot}},
  note         = {AI coding assistant}
}

@misc{cursor,
  author       = {{Anysphere}},
  title        = {Cursor},
  year         = {2026},
  howpublished = {\url{https://cursor.com/}},
  note         = {AI code editor}
}

@misc{windsurf,
  author       = {{Windsurf}},
  title        = {Windsurf},
  year         = {2026},
  howpublished = {\url{https://windsurf.com/}},
  note         = {AI code editor}
}

@book{Patton1990Qualitative,
  author    = {Patton, Michael Quinn},
  title     = {Qualitative Evaluation and Research Methods},
  year      = {1990},
  publisher = {SAGE Publications},
  address   = {Newbury Park, CA}
}

@book{Beyer1998Contextual,
  author = {Beyer, Hugh and Holtzblatt, Karen},
  title = {Contextual Design: Defining Customer-Centered Systems},
  year = {1997},
  isbn = {9780080503042},
  publisher = {Morgan Kaufmann Publishers Inc.},
  address = {San Francisco, CA, USA},
}

@misc{Hong2026SkillSmells,
      title={From Anatomy to Smells: An Empirical Study of SKILL.md in Agent Skills},
      author={David Boram Hong and Aaron Imani and Iftekhar Ahmed},
      year={2026},
      eprint={2607.01456},
      archivePrefix={arXiv},
      primaryClass={cs.SE},
      url={https://doi.org/10.48550/arXiv.2607.01456},
      doi={10.48550/arXiv.2607.01456}
}

@inproceedings{Yang2024SWEagent,
    author    = {Yang, John and Jimenez, Carlos E. and Wettig, Alexander and Lieret, Kilian and Yao, Shunyu and Narasimhan, Karthik and Press, Ofir},
    title     = {SWE-agent: agent-computer interfaces enable automated software engineering},
    year      = {2024},
    isbn      = {9798331314385},
    doi = {10.52202/079017-1601},
    url = {https://doi.org/10.52202/079017-1601},
    publisher = {Curran Associates Inc.},
    address   = {Red Hook, NY, USA},
    booktitle = {Advances in Neural Information Processing Systems 37},
    volume    = {37},
    pages     = {50528--50652},
    articleno = {1601},
    numpages  = {125},
    location  = {Vancouver, BC, Canada},
    month     = {December},
    note      = {Held 10--15 December 2024},
    series    = {NIPS '24}
}

@article{Cohen1960Kappa,
  author = {Cohen, Jacob},
  title = {A Coefficient of Agreement for Nominal Scales},
  journal = {Educational and Psychological Measurement},
  volume = {20},
  number = {1},
  pages = {37--46},
  year = {1960},
  doi = {10.1177/001316446002000104},
  url = {https://doi.org/10.1177/001316446002000104},
  publisher = {SAGE Publications}
}

@inproceedings{Zhang2022DietCode,
 author = {Zhang, Zhaowei and Zhang, Hongyu and Shen, Beijun and Gu, Xiaodong},
 title = {Diet code is healthy: simplifying programs for pre-trained models of code},
 year = {2022},
 isbn = {9781450394130},
 publisher = {Association for Computing Machinery},
 booktitle = {Proceedings of the 30th ACM Joint European Software Engineering Conference and Symposium on the Foundations of Software Engineering},
 address = {New York, NY, USA},
 url = {https://doi.org/10.1145/3540250.3549094},
 doi = {10.1145/3540250.3549094},
 pages = {1073–1084},
 numpages = {12},
 location = {Singapore, Singapore},
 series = {ESEC/FSE 2022}
}

@article{Xiao2026TrajectoryReduction,
 author = {Xiao, Yuan-An and Gao, Pengfei and Peng, Chao and Xiong, Yingfei},
 title = {Reducing Cost of LLM Agents with Trajectory Reduction},
 year = {2026},
 issue_date = {July 2026},
 publisher = {Association for Computing Machinery},
 address = {New York, NY, USA},
 volume = {3},
 number = {FSE},
 url = {https://doi.org/10.1145/3797084},
 doi = {10.1145/3797084},
 journal = {Proc. ACM Softw. Eng.},
 month = jun,
 articleno = {FSE056},
 numpages = {23}
}

@article{Hassan2026SE3,
 author = {Hassan, Ahmed E. and Oliva, Gustavo A. and Lin, Dayi and Chen, Boyuan and Jiang, Zhen Ming (Jack)},
 title = {Towards AI-Native Software Engineering (SE 3.0): A Vision and a Challenge Roadmap},
 year = {2026},
 publisher = {Association for Computing Machinery},
 address = {New York, NY, USA},
 issn = {1049-331X},
 url = {https://doi.org/10.1145/3807901},
 doi = {10.1145/3807901},
 note = {Just Accepted},
 journal = {ACM Trans. Softw. Eng. Methodol.},
 month = apr
}

@article{Ehsani2026TechnicalDebt,
 author = {Ehsani, Ramtin and Rawal, Shriya and Cai, Yuanfang and Chatterjee, Preetha},
 title = {Faster Code, Deeper Debt? A Multivocal Literature Review on Technical Debt and Its Early Signs in LLM-Assisted Software Development},
 year = {2026},
 publisher = {Association for Computing Machinery},
 address = {New York, NY, USA},
 issn = {1049-331X},
 url = {https://doi.org/10.1145/3820165},
 doi = {10.1145/3820165},
 note = {Just Accepted},
 journal = {ACM Trans. Softw. Eng. Methodol.},
 month = jun
}

@inproceedings{Reusens2024NativeBias,
  title = "Native Design Bias: Studying the Impact of {E}nglish Nativeness on Language Model Performance",
  author = "Reusens, Manon  and Borchert, Philipp  and De Weerdt, Jochen  and Baesens, Bart",
  editor = "Inui, Kentaro  and Sakti, Sakriani  and Wang, Haofen  and Wong, Derek F.  and Bhattacharyya, Pushpak  and Banerjee, Biplab  and Ekbal, Asif  and Chakraborty, Tanmoy  and Singh, Dhirendra Pratap",
  booktitle = "Proceedings of the 14th International Joint Conference on Natural Language Processing and the 4th Conference of the Asia-Pacific Chapter of the Association for Computational Linguistics",
  month = dec,
  year = "2025",
  address = "Mumbai, India",
  publisher = "The Asian Federation of Natural Language Processing and The Association for Computational Linguistics",
  url = "https://aclanthology.org/2025.findings-ijcnlp.73/",
  doi = "10.18653/v1/2025.findings-ijcnlp.73",
  pages = "1195--1215",
  ISBN = "979-8-89176-303-6",
}

@inproceedings{Sultana2026Bangladeshs,
  author = {Sultana, Sharifa and Samad, Rupali Tasnim and Haque, Mehzabin and Sultana, Zinnat and Jahangir, Zulkarin and Hossain, B M Mainul and Noman, Rashed Mujib and Ahmed, Syed Ishtiaque},
  title = {Bangladesh AI Readiness: Gaps in Curriculum, Infrastructure, and Governance},
  year = {2026},
  isbn = {9798400727023},
  publisher = {Association for Computing Machinery},
  address = {New York, NY, USA},
  url = {https://doi.org/10.1145/3811242.3819093},
  doi = {10.1145/3811242.3819093},
  booktitle = {Proceedings of the 2026 ACM SIGCAS/SIGCHI Conference on Computing and Sustainable Societies},
  pages = {433–446},
  numpages = {14},
  series = {COMPASS '26}
}

@article{Fetters2013Integration,
  author  = {Fetters, Michael D. and Curry, Leslie A. and Creswell, John W.},
  title   = {Achieving Integration in Mixed Methods Designs---Principles and Practices},
  journal = {Health Services Research},
  volume  = {48},
  number  = {6pt2},
  pages   = {2134--2156},
  year    = {2013},
  doi     = {10.1111/1475-6773.12117},
  url     = {https://doi.org/10.1111/1475-6773.12117}
}

@article{OCathain2010Three,
  author  = {O'Cathain, Alicia and Murphy, Elizabeth and Nicholl, Jon},
  title   = {Three Techniques for Integrating Data in Mixed Methods Studies},
  journal = {BMJ},
  volume  = {341},
  pages   = {c4587},
  year    = {2010},
  doi     = {10.1136/bmj.c4587},
  url     = {https://doi.org/10.1136/bmj.c4587}
}

\end{document}